\documentclass[useAMS,usenatbib]{mn2e}

\usepackage{aas_macros}
\usepackage{graphicx}
\usepackage{amssymb}
\usepackage{amsmath}
\usepackage{color}
\usepackage{changes}
\usepackage{multirow}

\newcommand{\Ms} {$\rm{M_{\odot}}~$}

\begin{document}

\title[Shape and Spin of Minihaloes II]
{Shape and Spin of Minihaloes. II: The Effect of Streaming Velocities} 
\author[Druschke et al.]{Maik Druschke$^{1}$\thanks{E-mail: 	
vu412@ix.urz.uni-heidelberg.de},
Anna T. P. Schauer$^{1,2}$\thanks{Hubble Fellow}, 
Simon C. O. Glover$^{1}$, Ralf S. Klessen$^{1}$\\
$^{1}$ Universit\"at Heidelberg, Zentrum f\"ur Astronomie, Institut f\"ur Theoretische
Astrophysik, Albert-Ueberle-Str. 2, 69120 Heidelberg, Germany\\
$^{2}$ Department of Astronomy, The University of Texas at Austin, Austin, TX 78712, USA}

\pagerange{\pageref{firstpage}--\pageref{lastpage}} \pubyear{2018}

\maketitle

\label{firstpage}

\begin{abstract}
Models of the decoupling of baryons and photons during the recombination epoch predict the existence of a large-scale velocity offset between baryons and dark matter at later times, the so-called streaming velocity. In this paper, we use high resolution numerical simulations to investigate the impact of this streaming velocity on the spin and shape distributions of high-redshift minihalos, the formation sites of the earliest generation of stars. We find that the presence of a streaming velocity has a negligible effect on the spin and shape of the dark matter component of the minihalos. However, it strongly affects the behaviour of the gas component. The most probable spin parameter increases from $\sim$0.03 in the absence of streaming to $\sim$0.15 for a run with a streaming velocity of three times $\sigma_{\rm rms}$, corresponding to 1.4 km\,s$^{-1}{}$ at redshift $z=15$. The gas within the minihalos becomes increasingly less spherical and more oblate as the streaming velocity increases, with dense clumps being found at larger distances from the halo centre. The impact of the streaming velocity is also  mass-dependent: less massive objects are influenced more strongly, on account of their shallower potential wells. The number of halos in which gas cooling and runaway gravitational collapse occurs decreases substantially as the streaming velocity increases. However, the spin and shape distributions of gas that does manage to cool and collapse are insensitive to the value of the streaming velocity and we therefore do not expect the properties of the stars that formed from this collapsed gas to depend on the value of the streaming velocity. The spin and shape of this central gas clump are uncorrelated with the same properties measured on the scale of the halo as a whole.

\end{abstract}

\section{Introduction}
The currently most favoured cosmological model for describing the evolution of our Universe since the Big Bang is the so-called $\Lambda$CDM (Lambda Cold Dark Matter) model. This is the simplest model that is consistent with existing cosmological measurements and is thus considered the standard model of Big Bang cosmology. In this model, most of the matter is dark and interacts with baryonic matter solely through gravity.
$\Lambda$CDM is a hierarchical model in which structure forms first on the smallest scales, with larger bound structures forming later via mergers and accretion. 

The first stars to form in the Universe -- the so-called Population III (Pop.\ III) -- therefore form in small-scale, low-mass bound structures known as dark matter minihaloes \citep[see e.g.][]{tegmark97,bcl99,abn00,yahs03}. Cooling in these systems takes place primarily via molecular hydrogen emission, with the result that the gas temperature never decreases below a few hundred Kelvin. The relatively high gas temperature inhibits fragmentation of the gas and promotes rapid gas accretion onto any protostars that form. Consequently, the initial mass function (IMF) of Pop.\ III stars is expected to be dominated by massive stars \citep[see e.g.][]{abn02,on07,tm08,y08,Hirano}, although it remains uncertain whether the IMF also extends down to values below a solar mass due to disk fragmentation \citep[see e.g.][]{Clark11,get11,get12,sb14,hb17,susa19}. Understanding the properties of the dark matter and gas making up the minihaloes that host the first generation of stars is of great importance for our understanding of the onset and outcome of Pop.\ III star formation.

In an earlier paper (\citealt{Druschke}; hereafter, Paper I), we used a high resolution cosmological simulation to investigate the distribution of shapes and spins of a large sample of dark matter minihaloes. In the subset of minihaloes in which the gas cools and undergoes runaway gravitational collapse (i.e.\ the minihaloes in which Pop.\ III star formation occurs), we also investigated whether the spin of the dense, cooling gas was correlated with the spin of the halo on large scales, and showed that there was no significant correlation. 

However, the simulation analyzed in Paper I started from initial conditions in which there was no relative velocity or ``streaming velocity'' between the dark matter and the gas. In reality we do not expect this to be the case. 
Before recombination, baryons and photons were tightly coupled by Compton scattering and behaved as if they were a single fluid. On the other hand, the dark matter was not directly coupled to this baryon-photon fluid and interacted with it only via gravity. Consequently, even though fluctuations in the density of the baryon-photon fluid and the dark matter are correlated, one nevertheless expects some motion of the fluid relative to the dark matter. In an influential paper, \citet{Tseliakhovich} pointed out that an imprint of this motion survives in the baryons even after recombination, in the form of a large-scale streaming motion of the baryons with respect to the dark matter. The resulting relative velocity (distributed like a multivariate Gaussian, with a standard deviation $\sigma \approx 30$\,km\,s$^{-1}$ at recombination) is highly supersonic shortly after recombination, but decays with the expansion of the Universe. It therefore plays little role in the formation of galaxies at low redshifts. However, several authors have shown that at high redshift, streaming velocities have a significant effect on the creation and formation of dark matter minihaloes and so-called atomic cooling haloes.\footnote{Dark matter haloes with virial temperatures $T \sim 10^{4} \: {\rm K}$ or above which are cooled by Lyman-$\alpha$ emission from atomic hydrogen.}
Amongst other effects, a non-zero streaming velocity suppresses the formation of minihaloes \citep{Tseliakhovich,naoz12} and increases the minimum halo mass required for efficient gas cooling and consequent Pop.\ III star formation \citep{Dalal,greif11,stacy11a,schauer18}
This leads to a delayed onset of Population~III star formation \citep{greif11,maio11}.
Other authors have studied the influence of streaming velocities on the shape of minihalos before \citep{Chiou}. Due to our implementation of H$_2{}$ cooling, the most important coolant at high redshift, we are able to follow the gas to high densities and investigate the proto-star forming regions in the halos.

In this paper, we therefore extend the analysis from Paper I to the case where the initial streaming velocity of the baryons relative to the dark matter is non-zero. We analyze several high-resolution simulations carried out with different streaming velocities and explore whether the value of the streaming velocity affects the spin or the shape distributions of the gas or the dark matter in minihaloes. The paper is structured as follows. In Section \ref{Simulation}, we give a short overview of the set of simulations from \cite{schauer18} that we use for our analysis, and in Section \ref{AnalysisMethods} we present our analysis methods. In Section \ref{Plots}, we look in detail at the changes that occur in a representative halo as we increase the streaming velocity, while in Section \ref{Results} we present the results of our analysis of the full set of haloes and provide the reader with a toy model. 
Finally we summarize our findings and results in Section \ref{Conclusion}.

\section{Simulation}\label{Simulation}
The cosmological simulations that we use for our analysis were previously described in \citet{anna17b} and \citet{schauer18}, and full details can be found in those papers. We therefore give here only a brief overview of the most important properties of the simulations.

The simulations were carried out using the moving mesh code {\sc arepo} \citep{arepo} and include both gas and dark matter. The chemical and thermal evolution of the gas were treated using a primordial chemistry network and cooling function based on the one presented in \citet{cgkb11}, but updated as described in \citet{schauer18}. Here, we choose to work with the four simulations from \citet{schauer18} that have a box size of 1\,cMpc/$h$, where the `c' denotes comoving units and $h$ is the value of the Hubble parameter in units of $100 \, {\rm km \, s^{-1} \, Mpc^{-1}}$. These simulations have a particle mass of
$\sim 100$\,\Ms\,for dark matter and a target mass of $\sim 20$\,\Ms\,for the gas cells\footnote{{\sc arepo} refines or de-refines gas cells as required to ensure that their masses stay within a factor of two of the specified target mass; see \citet{arepo} for more details.}. 
After creating the initial conditions with MUSIC \citep{hahn11} at $z=200$ with Planck parameters 
(namely $h=0.6774$, $\Omega_0 = 0.3089$, $\Omega_b = 0.04864$, $\Omega_\Lambda = 0.6911$, $n=0.96{}$ and $\sigma_8 = 0.8159$, \citealt{planck15}), the dark matter and gas are followed to redshift $z=14$. 
Even though these velocities are supersonic at the highest redshift, the effect they have on gas previous to $z=200{}$ are negligible \citep{park20}. 
In Paper I, we found that the minihalo spin and shape distributions evolve only weakly with redshift, and so in this paper we focus on the properties of the haloes at the final output time.

In order to mimic different regions of the Universe with different streaming velocities, a constant offset velocity term was added to the initial conditions. This offset velocity was arbitrarily chosen to point in the positive $x$ direction, but our results are independent of this choice, owing to the large-scale isotropy of the Universe. For the four different boxes, the amplitude of the streaming velocity was set to 0, 1, 2 and 3 times $\sigma_\mathrm{rms}$, corresponding to 0, 6, 12 and 18\,km\,s$^{-1}{}$ at $z=200$. The no streaming run was previously analyzed in Paper I but is included here for the purposes of comparison.

\section{Analysis}\label{AnalysisMethods}
This paper focuses primarily on the spin and the shape of  
minihaloes at different streaming velocities. Before we discuss our results, we introduce 
some important physical quantities. A more detailed description can be found in Paper I.

We start with the angular momentum $\vec{J}(R){}$, which is defined for each minihalo by summing up the values of every dark matter particle or gas cell within a distance $R$ from the most bound cell, which we take to define the centre of the halo:
\begin{equation}
\begin{split}
 \vec{J}(R) = \sum\limits_{r_{i}<R} m_{i} \vec{r}_{i} \times \vec{v}_{i}.
\end{split}
\end{equation}
Here, $m_i$ is the mass of the particle or gas cell, $\vec{r}_i$ the distance from the particle or gas cell to the center of the halo and $\vec{v}_i$ the velocity relative to the center of the halo. In the special case where $R = R_{\rm vir}$ (the virial radius), this equation yields the angular momentum of the halo as a whole. 

Additionally, we compute the inertia tensor in order to calculate the side lengths of a halo 
\citep{Springel2004}:
\begin{equation}
I_{jk} = \sum_{i=1}^N m_{i}(|\vec{r}_{i}|^{2}\delta_{jk}-r_{i,j}r_{i,k}) .
\end{equation}  
Here, $\delta_{jk} $ represents the Kronecker delta, $m_{i}{}$ is the mass of the $i$-th gas cell or dark matter particle, 
$\vec{r}_{i}$ is its distance from the halo centre, and $r_{i,j}$ and $r_{i,k}$ and the $j$-th and $k$-th components of $\vec{r}_{i}$. 
The centre of the halo is defined by the position of the most bound dark matter particle, equivalent to the potential minimum of the halo. 
Using the eigenvalues $I_{1}$, $I_{2}$, $I_{3}$ of this tensor and the halo mass, the side lengths $a \geq b \geq c{}$ of an ellipsoid can be calculated (see Paper I), which can then be used to compute the sphericity and triaxiality of the halo, as outlined in Section~\ref{sec:shape} below.

\subsection{Spin}
We are interested in how fast a halo is rotating, independent of its mass. 
We therefore choose to work with the spin parameter $\lambda^{\prime}$, and follow the definition of \cite{Bullock}:
\begin{equation}\label{LambdaGasDM}
\lambda^{\prime}_{c}(R)= \dfrac{\vert J \vert_{c}(R)}{\sqrt{2} R\, M_{c}(R)V_{\mathrm{circ}}(R)}. 
\end{equation}
As before, $J_{c}$ describes the angular momentum and $M_{c}$ the mass. The index $c{}$ stands for the components (gas, dark matter, or the total matter content of the halo) that we are interested in. Furthermore, $R$ is the radius up to which the particles or gas cells are considered and $V_{\mathrm{circ}}$ is the circular velocity which is given by $V_{\mathrm{circ}}(R)^{2} =R^{-1} G M(R) $.

The spin parameter can take values between $\lambda^{\prime} = 0$ and $\lambda^{\prime} = 1$. 
A spin parameter of 0 corresponds to a non-rotating halo, 
while a spin parameter of 1 corresponds to Keplerian rotation of all particles and gas cells. 

Statistically, the spin of a set of haloes can be approximated by a log-normal distribution \citep{Warren,Mo}:
\begin{equation}
P(\lambda^{\prime})  = \dfrac{1}{\lambda^{\prime} \sqrt{2\pi}\sigma_{0}} \mathrm{exp\left(-\dfrac{ln^{2}\left(\frac{\lambda^{\prime}}{\lambda_{0}}\right)}{2\sigma_{0}^{2}}\right)} .
\end{equation}\label{Lambda}
The most probable value of this well-known log-normal distribution is referred as the peak value in the following. 
In regions of the Universe without streaming velocity,
previous studies (\citealt{Sasaki}, Paper I) find peak values of $\lambda^{\prime} \simeq 0.03$. The distribution of the spin parameter of all $\sim8000{}$ halos  with a mass of at least $\mathrm{M_{min}} = 10^{5}\ \mathrm{M_{\odot}}$ (corresponding to roughly 1000 gas and dark matter particles) at $z=14$ can be seen in Figure 1 (left panel),  The distribution shows a peak value of $\lambda^\prime=0.0294$. 
\begin{figure} 
\includegraphics[width=0.99\columnwidth]{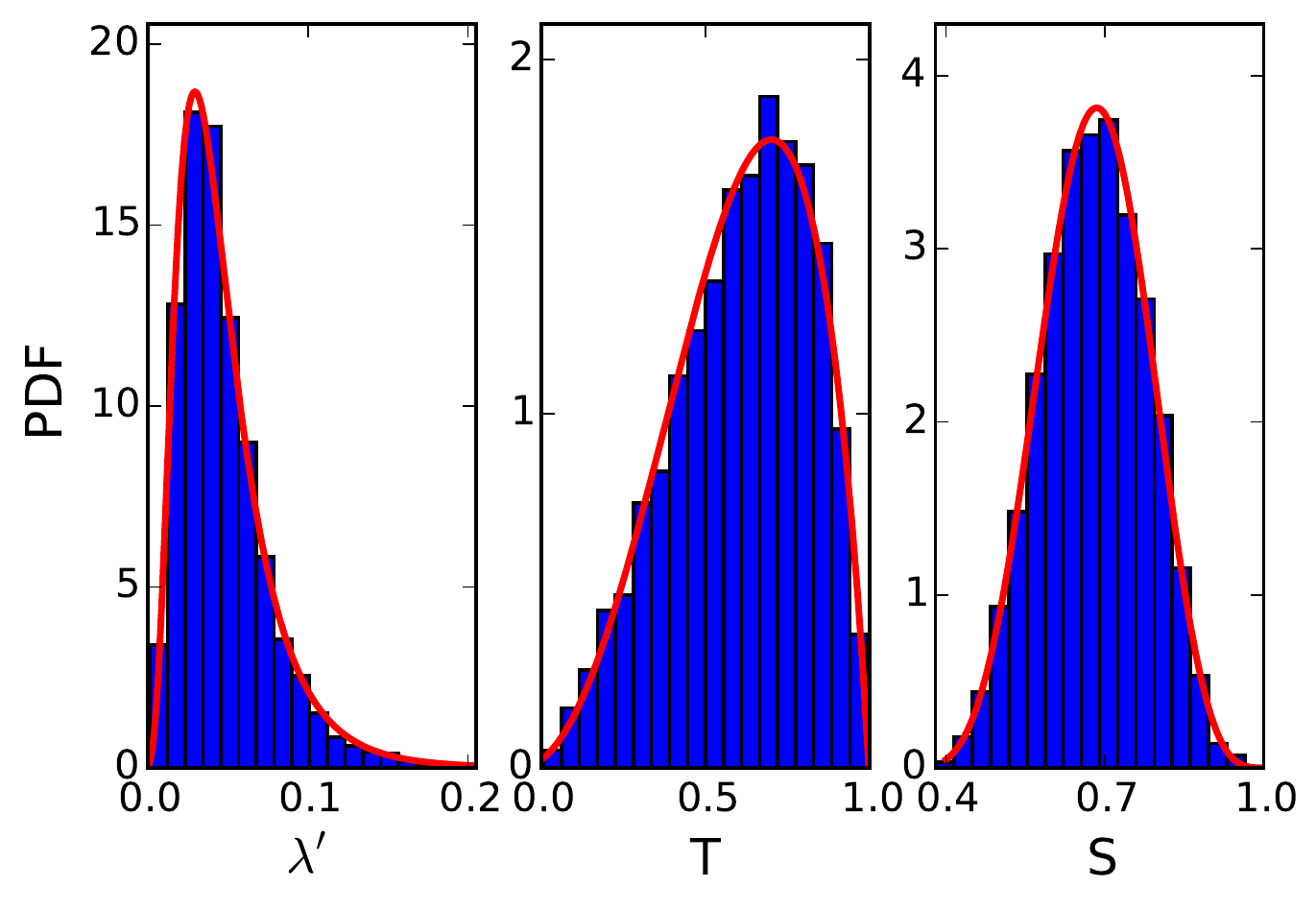}
\caption{Distribution of the halo properties for all $\sim8000{}$ haloes with $M \geq M_{\rm min}$  in a simulation without streaming velocity 
at redshift $z=14$. Left panel: spin parameter. The red curve is a log-normal fit to the data, with parameters as given in the text. Middle panel: triaxiality. The red curve is a beta-distribution fit to the data, with parameters as given in the text. Right panel: sphericity. Again, the red curve is a beta-distribution fit to the data, with parameters given in the text.}
\label{fig:LambdaTriaxSphaereVsPDF}
\end{figure}
\subsection{Shape}
\label{sec:shape}

In order to quantify the shape of the haloes, we follow the definition of \cite{Springel2004}. The triaxiality and sphericity can be determined using the side lengths $a \geq b \geq c{}$ of the ellipsoid that we determined above using the inertia tensor.

\subsubsection{Triaxiality}\label{Triaxiality}
The triaxiality can be calculated as follows \citep{Franx}:
\begin{equation}
T = \dfrac{a^{2}-b^{2}}{a^{2}-c^{2}} .  \label{triax-eq}
\end{equation}
This definition classifies haloes from oblate (low triaxiality,  $\mathrm{T}=0$) to prolate (high triaxiality, $\mathrm{T}=1$).
In the middle panel of Figure \ref{fig:LambdaTriaxSphaereVsPDF}, we show the triaxiality distribution of all haloes with masses $M > M_{\rm min}$ at redshift $z=14$ in the simulation without streaming velocities. It can be described by a beta distribution displayed in red (see also Paper I):
\begin{equation}
P(T)=\dfrac{\Gamma(a+b)T^{a-1}(1-T)^{b-1}}{\Gamma(a)\Gamma(b)}
\end{equation}
Since the gamma function $\Gamma$ serves only for normalization, the shape of the distribution is described by the two fit variables $a = 3.355$ and $b = 1.892$. The peak of the best fitting distribution is at $\mathrm{T_{peak}=0.700}$. 
Other studies \citep{Jang,Sasaki} have already shown that most haloes in our mass range are prolate, while haloes are more oblate at higher masses \citep[see e.g.][]{Warren,Allgood}.

\subsubsection{Sphericity}\label{sphericity}
The sphericity is defined as the ratio of the smallest to the largest side length of the halo ellipsoid: 
\begin{equation}
S = \dfrac{c}{a} .  \label{sphericity-eq}
\end{equation}
A sphericity of $S=1$ describes a perfectly spherical halo, while a lower value represents a stronger deviation from a perfect sphere. Analogously to the triaxiality, the sphericity distribution of all haloes is well described by a beta-distribution, which is shown in the right panel of Figure \ref{fig:LambdaTriaxSphaereVsPDF} for the simulation without streaming velocity at redshift $z=14$. Here, the shape parameters are $a=6.582$ and $b=5.750$ and the most probable value of $S$ is $S=0.685$.

The sphericity in the DM-only study by \cite{sasaki15} is slightly lower, at around $s \approx 0.45$, depending on halo mass and formation time. 
\citet{Chiou} study the shape of the minihaloes as well, with slightly varying 
definition of the parameters. Their prolateness factor $\xi = R_\mathrm{max} /R_\mathrm{min}{}$ is roughly the inverse of our sphericity, and peaks at 
$\xi \approx 1.3$
corresponding to $s \approx 0.77{}$ for no streaming velocity, and moves to higher values ($\xi \approx 1.6{}$ or $s \approx 0.63$) for a streaming velocity $v_\mathrm{bc} = 2\sigma_\mathrm{rms}$. This is in agreement with our findings of a gas sphericity of $s=0.744$ for no streaming and 
$s=0.637{}$ for $2\sigma_\mathrm{rms}$ streaming.

\section{Impact of streaming on individual haloes}\label{Plots}
%
\begin{figure*}
\includegraphics[width=1.99\columnwidth]{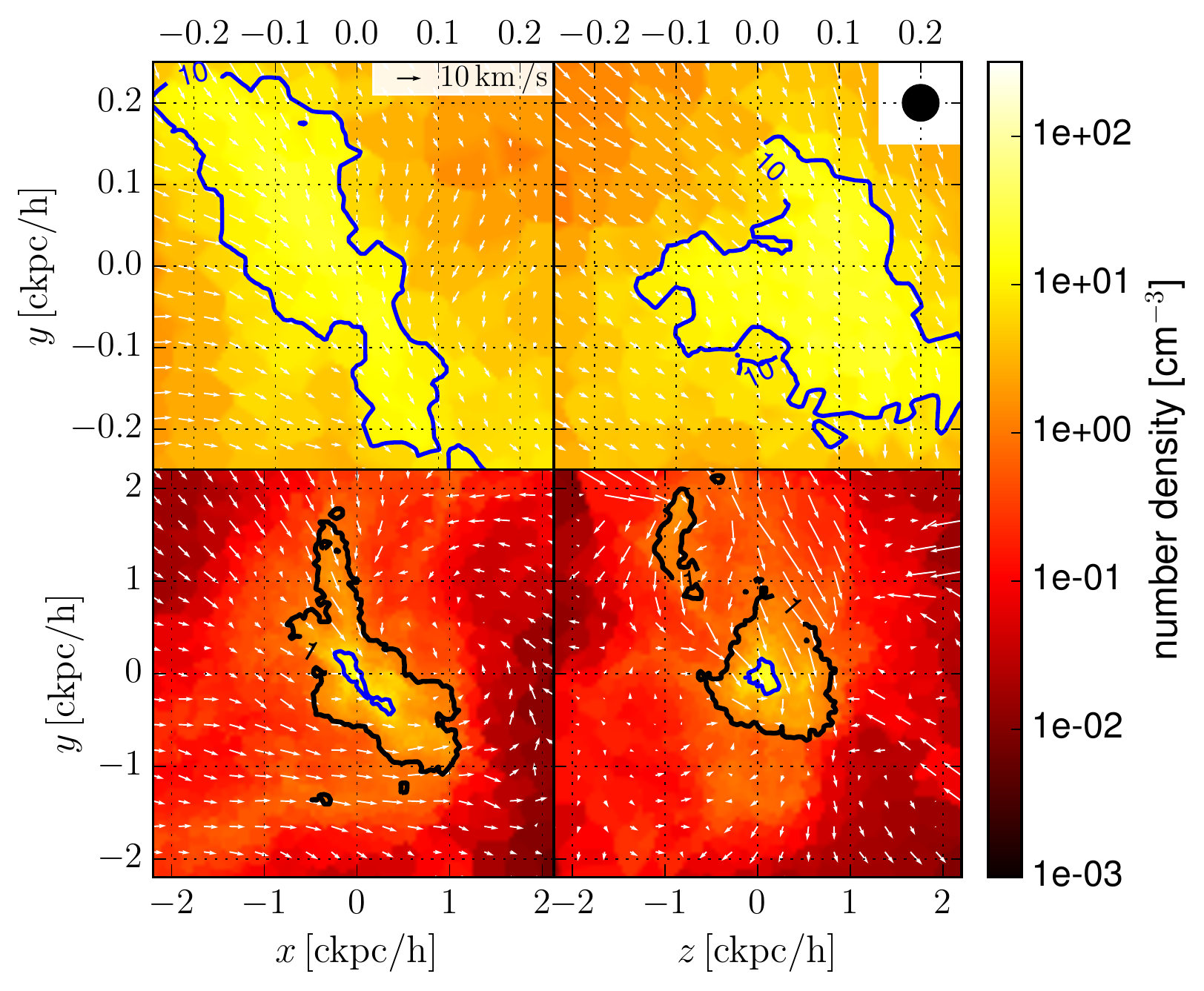}
\caption{Example of a typical minihalo with $\lambda^{\prime} = 0.0483$ and $M = 	9.57\times 10^{6}\, \mathrm{M}_\odot$ at $z = 14$ with 2$\sigma$ streaming. The left panels show slices through the minihalo in the $x-y$ plane, while the right panels show slices in the $z-y$ plane with respect to the halo center (x=0, y=0, z=0), which is defined by the most bound particle. The slices are color-coded by number density, and contours indicating number densities of 1, 10 and 100$\ \mathrm{cm}^{-3}$ are also shown. In the upper panels, the region within a box of side length $0.5$ ckpc/$h$ is shown, while in the lower panels a larger region of side length $3.0$ ckpc/$h$ is shown. The arrows indicate the direction and magnitude of the gas velocity.
For reference, we show a black arrow corresponding to a velocity of $10 \, {\rm km \, s^{-1}}$ in the top left panel. The dark circle in the upper right corner has a radius equal to the gravitational softening length.} 
\label{fig:SubPlot_Nr13_Final_0_25_1_5_hot}
\end{figure*}
Before examining the effect that a non-zero streaming velocity has on the full distribution of haloes, it is informative to study its effects in detail on a representative halo. In Figure \ref{fig:SubPlot_Nr13_Final_0_25_1_5_hot}, we show slices of density in the $x$-$y$ and $y$-$z$ planes through the centre of a halo (placed at the center of the figure) taken from the simulation with 2$\sigma$ streaming. This halo has a mass $M = 9.57\times 10^{6}\, \mathrm{M}_\odot$ and a spin parameter $\lambda^{\prime} = 0.0483$. 
For a better overview, the halo is shown at two different scales. In the upper panels, the slice has a side length of 0.5 ckpc/h, showing the inner core and thus the cold dense gas associated with the halo. In the lower panels, the side length is 3.0 ckpc/h, which allows a more general overview of the position and shape of the entire halo. The origin of the coordinate system is taken to be at the center of the halo (defined as the most bound particle), and the coordinate system is aligned so that the halo angular momentum vector points in the position $z$ direction. The colour-coding indicates the number density of the gas, ranging from 
$n_\mathrm{min}=0.001$\,cm$^{-3}{}$ (dark red) to $n_\mathrm{max}=250$\,cm$^{-3}{}$ (bright yellow). 
In addition, we have added contours for density thresholds of $n_\mathrm{thres}=1$\,cm$^{-3}{}$ (black), $10$\,cm$^{-3}{}$ (blue) and $100$\,cm$^{-3}{}$ (red). The white arrows indicate the direction of the gas velocity and their length indicates the magnitude of the velocity with respect to the centre of the halo. For comparison, the black arrow on a white background in the top of the image is normalized to a velocity of 10 km/s. Furthermore, the black circle on top right in the Figure indicates the radius of the gravitational softening length.

In comparison to Paper I, where similar plots are shown for a pair of haloes taken from a run with no streaming, we see that here, the center of rotation can deviate from the halo center. 
For example, the halo shown in Figure \ref{fig:SubPlot_Nr13_Final_0_25_1_5_hot} has a rotational 
centre at $x \approx y \approx$\,0.5\,ckpc/$h$, which can be seen in the lower left panel.

\begin{figure*}
\includegraphics[width=1.99\columnwidth]{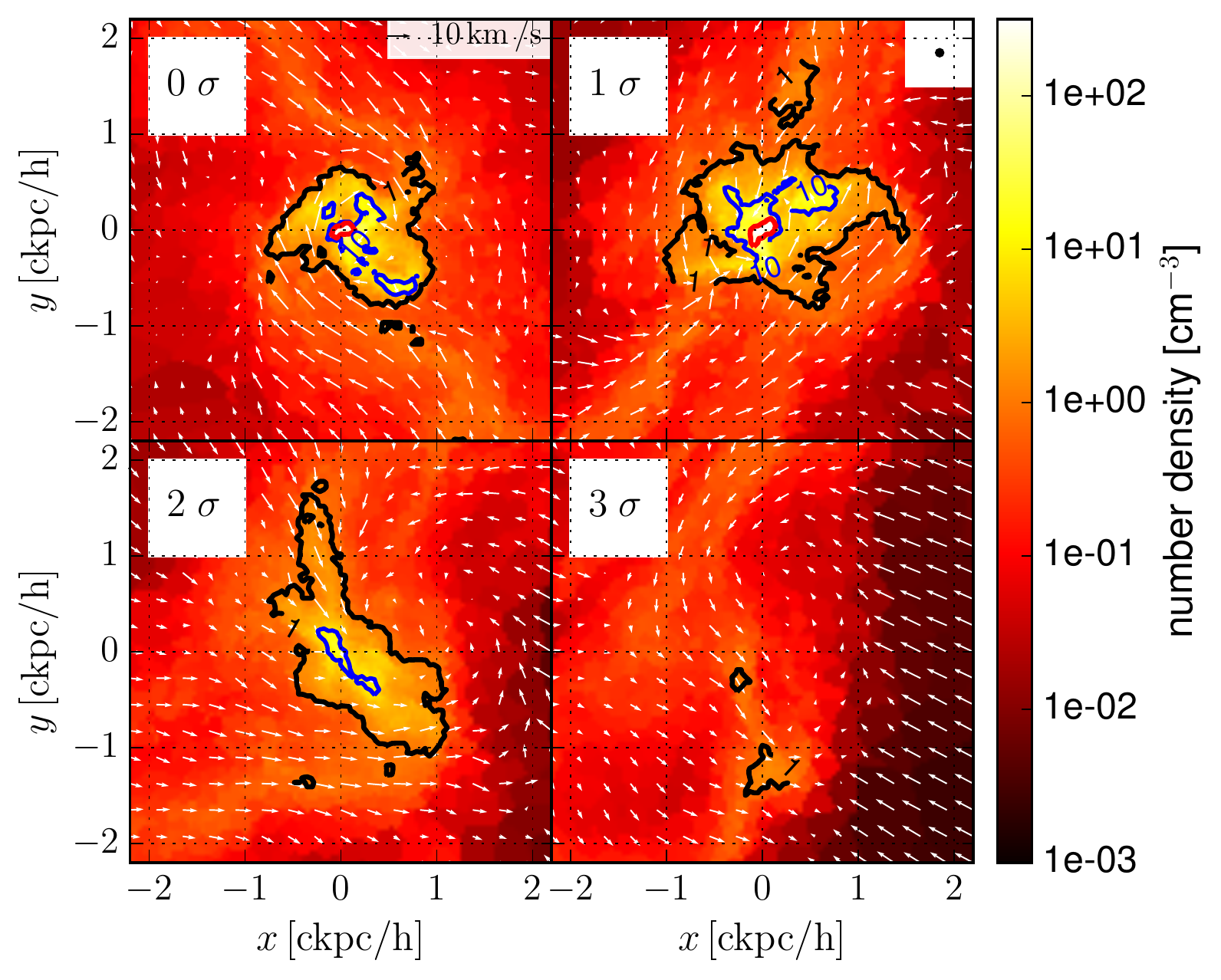}
\caption{As Figure~\ref{fig:SubPlot_Nr13_Final_0_25_1_5_hot}, 
but showing the same halo again at $z = 14$ in our four simulations with different streaming velocities. It is clear that as the streaming velocity increases, the amount of dense gas associated with the minihalo decreases.} 
\label{fig:PlotAlleSigmaNummer13}
\end{figure*}

In Figure \ref{fig:PlotAlleSigmaNummer13}, we show the same minihalo in all four simulations with different streaming velocities. Since we use the same initial conditions for all four simulations (other than the gas velocity offset), we can identify the same dark matter halo in all four simulations and compare its properties.

We see immediately that the higher the streaming velocity, the lower the maximum density in the minihalo. The core of the halo in the 0$\sigma$ run contains a lot of cold dense gas. Near the center, several clumps with number densities exceeding $50$\,cm$^{-3}{}$ have formed and the innermost clump contains gas at a density of more than $100$\,cm$^{-3}{}$. Even in the 1$\sigma$ streaming run, there is still a lot of dense gas visible. With 2$\sigma$ streaming, however, the halo contains no gas denser than $50$\,cm$^{-3}{}$, while at 3$\sigma$ the densest gas is $\sim 1 \, {\rm cm}^{-3}$, only a factor of a few larger than the mean halo density. 

As we will see later, the behaviour of this particular halo is quite typical. In the no streaming run, there are 206 haloes at $z=14$ that contain cold gas that is denser than $100 \, {\rm cm^{-3}}$, but in the 2$\sigma$ streaming run this number has dropped to 15 haloes, while in the 3$\sigma$ case we find only a single halo with cold gas above this density.

Figure~\ref{fig:PlotAlleSigmaNummer13} also illustrates another important phenomenon related to strong streaming, namely that the peak gas overdensity can become strongly offset from the halo center. There is a hint of this effect in the run with 2$\sigma$ streaming, but in this case the offset between the density peak and the halo center (which is always located at the origin) is small. In the case of 3$\sigma$ streaming, however, the offset between the highest gas densities and the halo center is clearly apparent. This offset can result in the formation of baryon-dominated clouds outside of the virial radius of the closest dark matter halo (\citealt{nn14,pnmv16,Hirano2018b}, see also \citealt{Chiou19} in the context of globular cluster formation). We discus this in detail with a toy model in Section \ref{sect:toy} . 

\section{Statistical analysis of the full minihalo sample}\label{Results}
As a next step, we study the distribution of spins and shapes in a statistical manner. For this part of our analysis, we selected all haloes with a (dark matter and gas combined) mass of at least $M_{\rm min} = 10^{5} \ \mathrm{M_{\odot}}{}$, as less massive haloes are not well resolved in our cosmological simulations (\citealt{schauer18}; Paper I). This mass limit leads to the selection of 
7982 haloes for no streaming, 6282 haloes for 1$\sigma$ streaming, 4798 haloes for 2$\sigma$ streaming and 4263 haloes for 3$\sigma$ streaming. The different number of haloes results from the delaying effect that streaming motions have on the formation and growth of minihaloes \citep[see e.g.][]{Tseliakhovich,greif11,mkc11,stacy11a}.
\subsection{Spin and shape distribution}\label{Spin and shape properties}
%
\begin{figure}
\includegraphics[width=0.99\columnwidth]{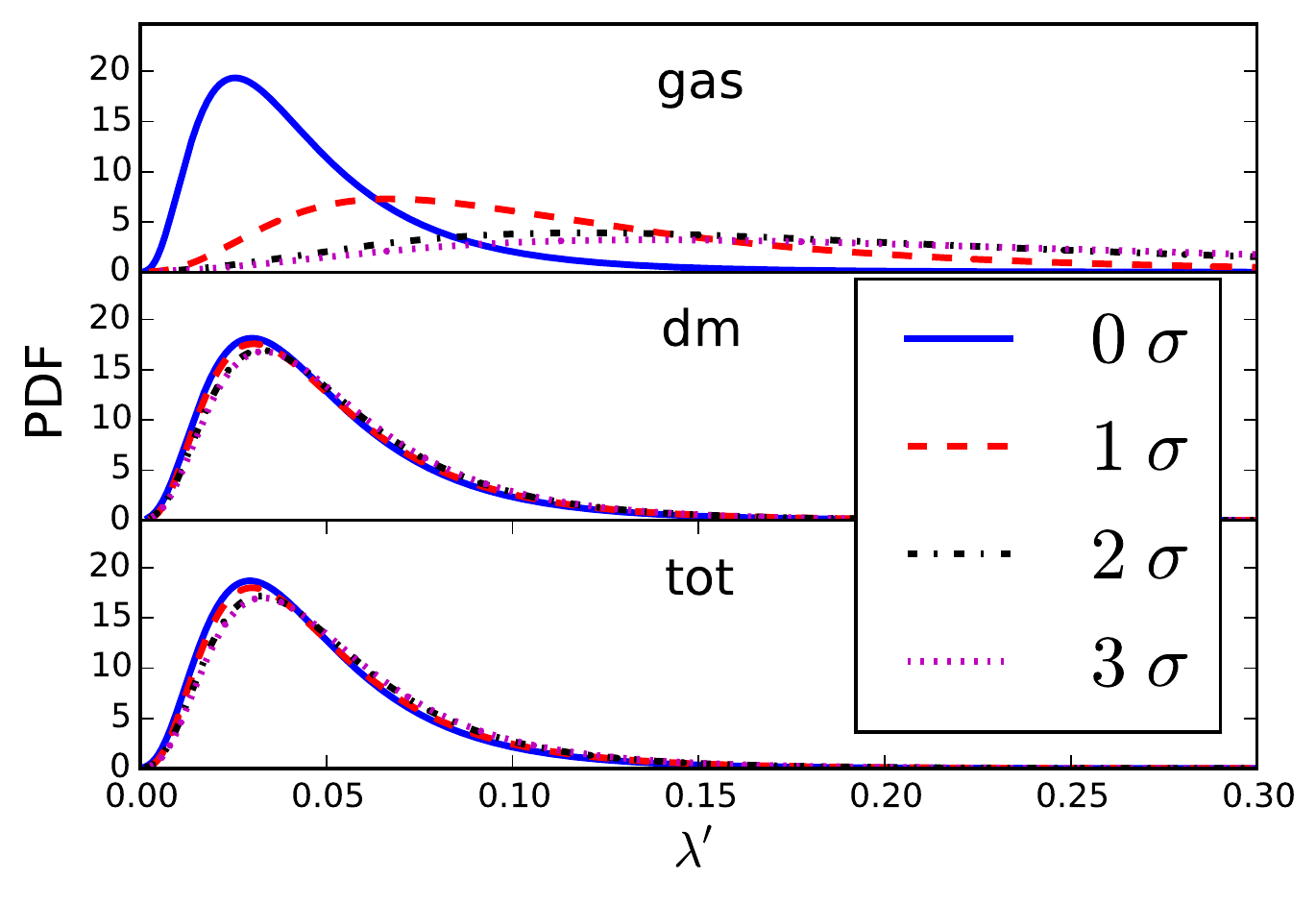}
\caption{The distributions of the spin parameters of all haloes for gas (upper panel), dark matter (middle panel) and total matter (lower panel). The blue solid ($0\sigma$), red dashed ($1\sigma$), black dash and dotted ($2\sigma$) and purple dotted ($3\sigma$) lines indicate the respective streaming velocities. While the spin parameter of the gas shifts to higher values for larger streaming velocity values, the dark matter and total spin distributions remain largely unchanged. } 
\label{fig:LambdaGASdmTOT0123Subplot}
\end{figure}
In Figure \ref{fig:LambdaGASdmTOT0123Subplot}, we show the spin parameter distribution for gas (top panel), dark matter (middle panel) and the total halo (bottom panel) for all four streaming velocity simulations.\footnote{To improve the visibility, we only show the fit of the log-normal distribution (compare Figure \ref{fig:LambdaTriaxSphaereVsPDF}), and not the entire histogram.} 
As we can see, the distribution for the dark matter does not change significantly as we vary the streaming velocity. The same is true for the spin parameter distribution for the halo as a whole, since this is dominated by the dark matter contribution. 
The peak value varies between $\lambda^{\prime}_\mathrm{dm} = 0.030$ and $\lambda^{\prime}_\mathrm{dm} = 0.033$ for the dark matter and between $\lambda^{\prime}_\mathrm{tot} = 0.029$ and $\lambda^{\prime}_\mathrm{tot} = 0.033$ for the total mass. 

For gas on the other hand, the distribution is strongly affected by the size of the streaming velocity. Increasing the streaming velocity flattens the spin parameter distribution and shifts its peak toward much higher values of $\lambda^{\prime}$: it changes from $\lambda^{\prime}_\mathrm{gas} = 0.026$ with no streaming to $\lambda^{\prime}_\mathrm{gas} = 0.138$ for 3$\sigma$ streaming. These results are broadly consistent with the recent study of \cite{Chiou}, who examine the minihalo spin parameter distribution in simulations with no stream and 2$\sigma$ streaming. They also find a significant shift in the peak of the distribution with increasing streaming velocity, with $\lambda^{\prime}_\mathrm{gas}$ increasing from 0.04 in the case without streaming at $z = 10$ to 0.12 in the case of 2$\sigma$ streaming. 

\begin{figure}
\includegraphics[width=0.99\columnwidth]{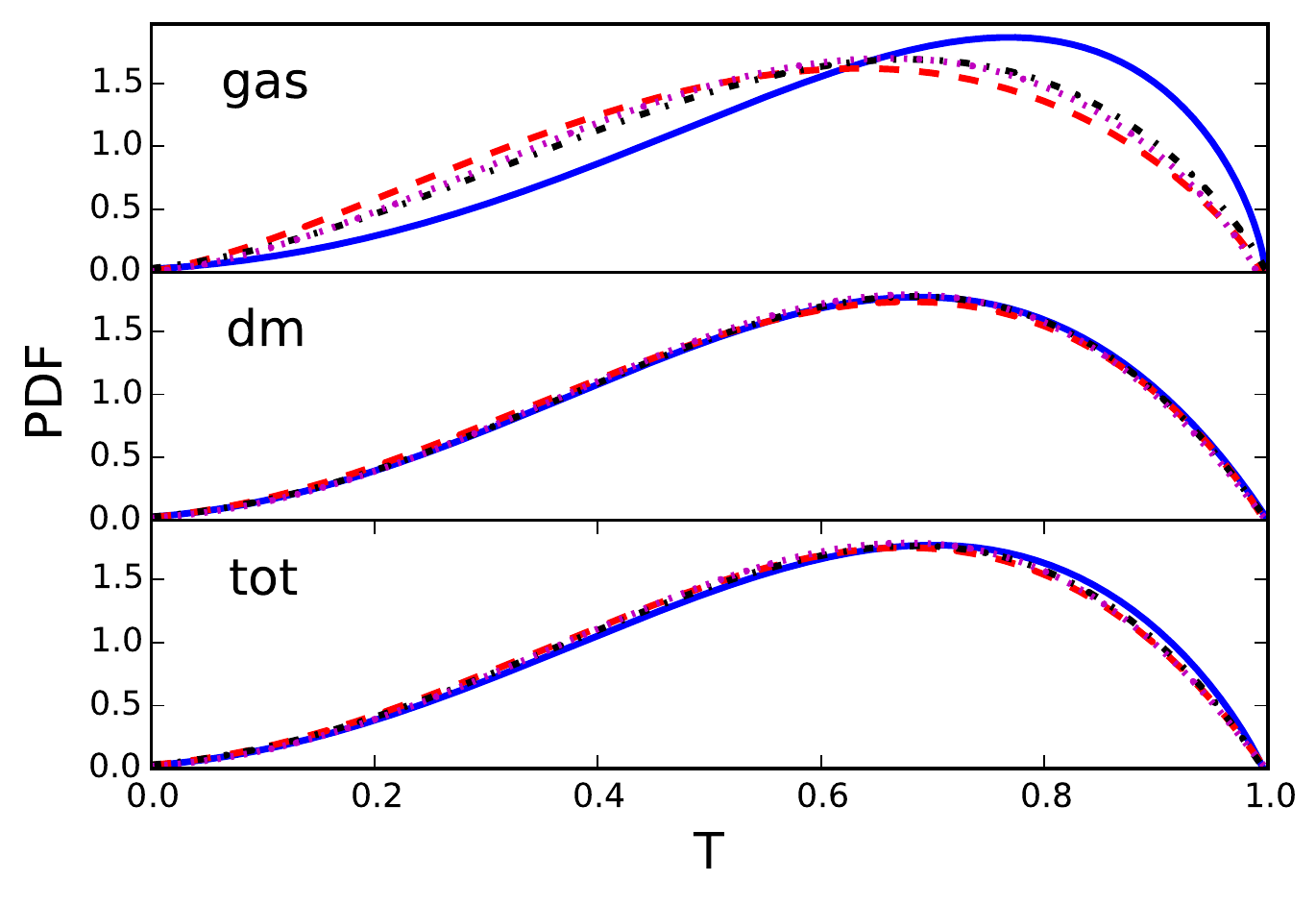}
\caption{Same as in Figure \ref{fig:LambdaGASdmTOT0123Subplot}, but for the triaxiality of all haloes. The triaxiality of the gas component is smaller for non-zero than for zero streaming velocity values, while the dark matter and total triaxiality distributions do not differ by much for these two cases. } 
\label{fig:TriaxGASdmTOT0123Subplot}
\end{figure}

\begin{figure}
\includegraphics[width=0.99\columnwidth]{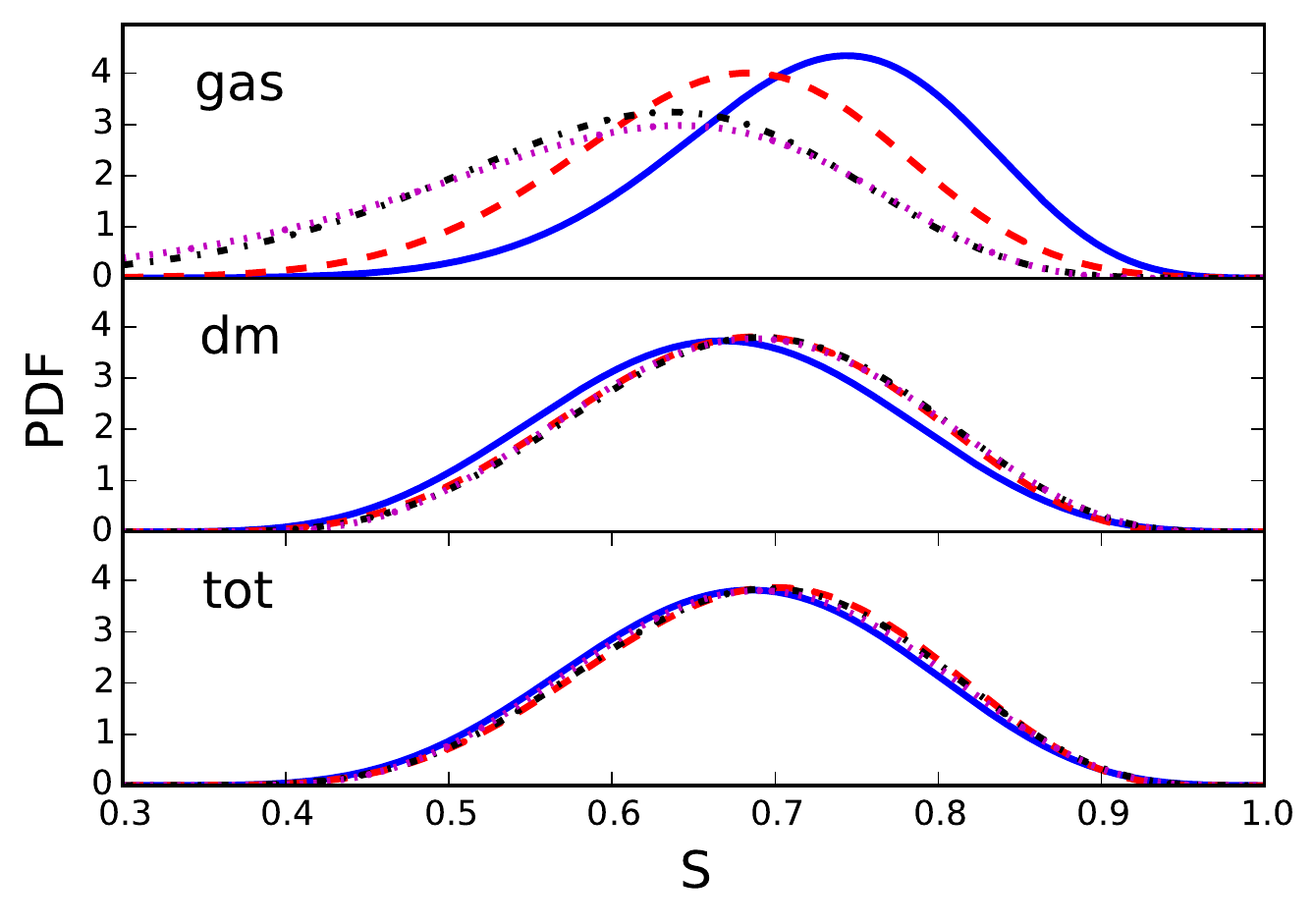}
\caption{Same as Figure \ref{fig:LambdaGASdmTOT0123Subplot}, but for the sphericity of all haloes. The sphericity of the gas component shifts to smaller values with larger streaming velocity values. The dark matter component and the total sphericity are, similiar to the spin and traixiality distributions, almost the same. } 
\label{fig:SpaereGASdmTOT0123Subplot}
\end{figure}

We have also examined how the shape of the minihaloes changes due to the streaming velocity. For this purpose, we show beta distribution fits to the triaxiality and sphericity distributions in Figures \ref{fig:TriaxGASdmTOT0123Subplot} and \ref{fig:SpaereGASdmTOT0123Subplot}, respectively. Again, we see that there are major changes in the distribution of gas, while the distribution of dark matter and total matter remains virtually unchanged. As the streaming velocity increases, both the sphericity and the triaxiality of the gas distribution decrease (i.e.\ it becomes less spherical and more oblate).
The peak values of $\lambda^{\prime}$, $S$ and $T$ at $z=14$ for each component in each simulation are listed in Table \ref{Table:ValuesTabelle}.

Finally, we have investigated whether there is any correlation between the shape and the spin parameter of the halos in any of our simulations. We find that in practice there does not appear to be any significant correlation between $\lambda^{\prime}$ and either $S$ or $T$, regardless of the value of the streaming velocity.

\begin{center}
\begin{table}
\centering
\begin{tabular}{|l|l|l|l|l|}
\hline
 \multicolumn{1}{|c|}{$\sigma_{\rm rms}$}          &     & \multicolumn{1}{c}{$\lambda^{\prime}$} & \multicolumn{1}{c}{T}   & \multicolumn{1}{c}{S}  \\ \hline
\multicolumn{1}{|c|}{\multirow{3}{*}{0}} & gas &   0.0256           &0.768&0.744  \\ \cline{2-5} 
\multicolumn{1}{|c|}{}                   & dm  &   0.0300           &0.688&0.668  \\ \cline{2-5} 
\multicolumn{1}{|c|}{}                   & tot &   0.0294			&0.700&0.685  \\ \hline
\multicolumn{1}{|c|}{\multirow{3}{*}{1}} & gas &   0.0672			&0.634&0.683  \\ \cline{2-5} 
                                         & dm  &   0.0305			&0.679&0.688  \\ \cline{2-5} 
                                         & tot &   0.0301			&0.676&0.700  \\ \hline
\multicolumn{1}{|c|}{\multirow{3}{*}{2}} & gas &   0.1193			&0.678&0.637  \\ \cline{2-5} 
                                         & dm  &   0.0326			&0.684&0.691  \\ \cline{2-5} 
                                         & tot &   0.0323			&0.683&0.696  \\ \hline
\multicolumn{1}{|c|}{\multirow{3}{*}{3}} & gas &   0.1381  			&0.661&0.642  \\ \cline{2-5} 
                                         & dm  &   0.0331  			&0.679&0.687  \\ \cline{2-5} 
                                         & tot &   0.0330			&0.677&0.691  \\ \hline
\end{tabular}
\caption{List of all calculated peak values at $z=14$ for the spin ($\lambda^{\prime}$), triaxiality (T) and sphericity (S) distributions, for our four simulations and for all components: gas, dark matter (dm) and total matter (tot).}
\label{Table:ValuesTabelle}
\end{table}
\end{center}
\subsubsection{Mass dependence} \label{Mass dependence}
\begin{figure}
\includegraphics[width=0.99\columnwidth]{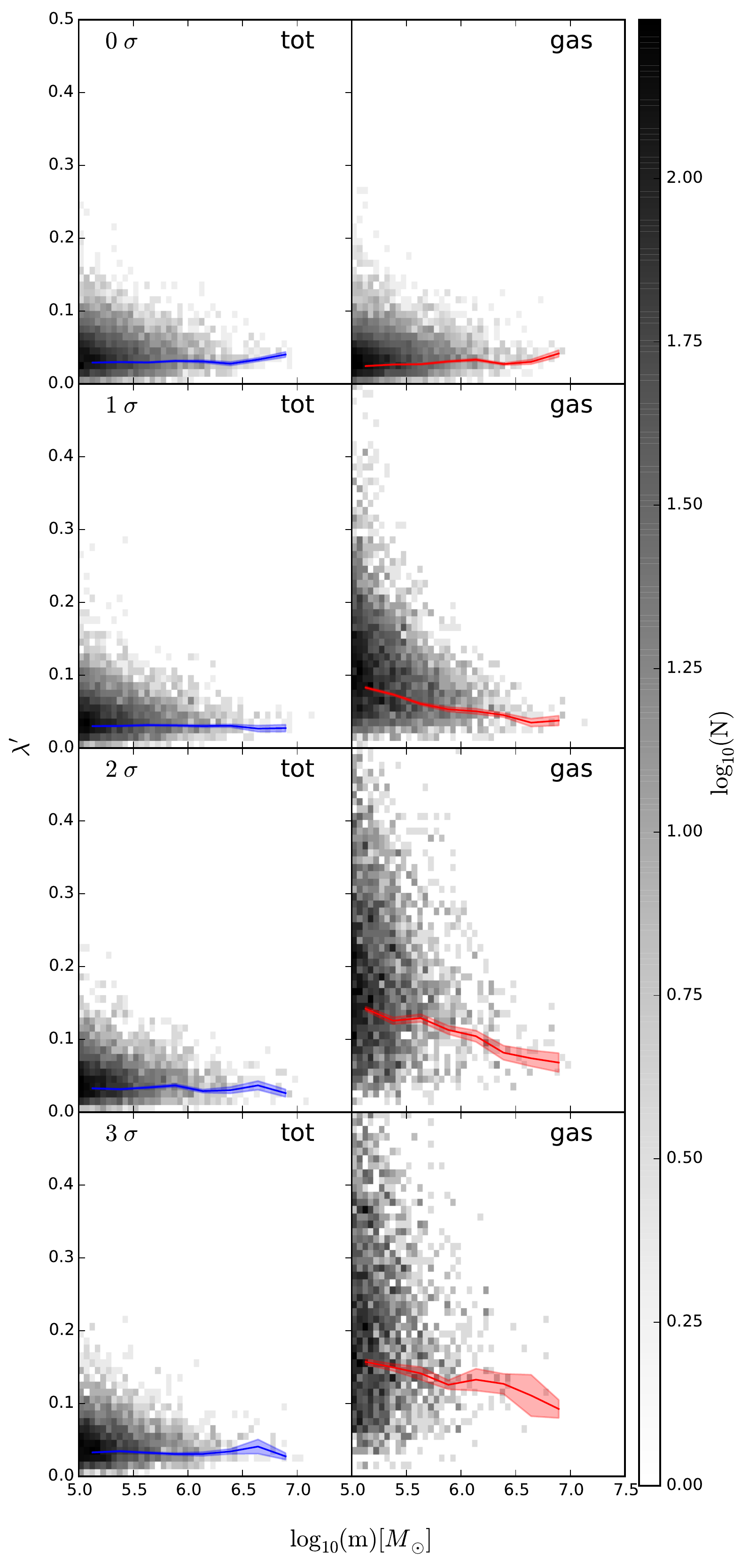}
\caption{Two dimensional histograms for the total spin parameter (left) and spin parameter of the gas component (right), as a function of the total mass of the minihalo. The solid lines show the peak value of the spin parameter distribution in that halo mass bin, and the shaded regions indicate the error in the determination of this peak value, estimated using the bootstrap method. We do not see a dependence of the spin parameter on halo mass for the total halo, or the zero streaming velocity case. For larger streaming velocities, the gas component of the halo increases, and does so most for lower mass haloes. }
\label{fig:Mass_vs_Lambda_0Sig_013}
\end{figure}

In order to understand why the spin and shape parameters for the gas change as the streaming velocity increases, we have explored how they vary as a function of the halo mass. In Figure \ref{fig:Mass_vs_Lambda_0Sig_013}, we show 2D histograms of the spin parameter as a function of the halo mass $M$ for the whole halo (left-hand panels) and for the gas component (right-hand panels) for each of the simulations. In these histograms, the spin is plotted against the logarithmically scaled mass using a $40 \times 40$ pixel grid. The colour of the pixels indicates the number of haloes contained in each.
We also show the most probable spin parameter (i.e.\ the peak in the distribution) for each halo mass (solid line). The shaded region around this line is an estimate of how accurately we are able to determine the peak in the distribution, computed using the bootstrap method.

We can clearly see that in the case of no streaming velocity, there is no correlation between the spin parameter and the halo mass, neither for the gas nor for the total halo. Instead, the peak value of the spin parameter distribution remains approximately constant with halo mass. Although there are haloes with relatively large values for the spin parameter ($\lambda^{\prime} > 0.15$), these are all found in the lowest mass bins with $M \sim 10^{5}\ \mathrm{M_{\odot}}$. Additionally, these haloes represent the high $\lambda^{\prime}$ tail of the distribution and may be missing in the higher mass bins simply because there are far fewer minihaloes in total present in those bins. We note that this is not a new result: other studies of the minihalo spin parameter distribution in the absence of streaming have also found it to be independent of halo mass \citep{Hirano,Sasaki}.

\begin{figure}
\includegraphics[width=0.99\columnwidth]{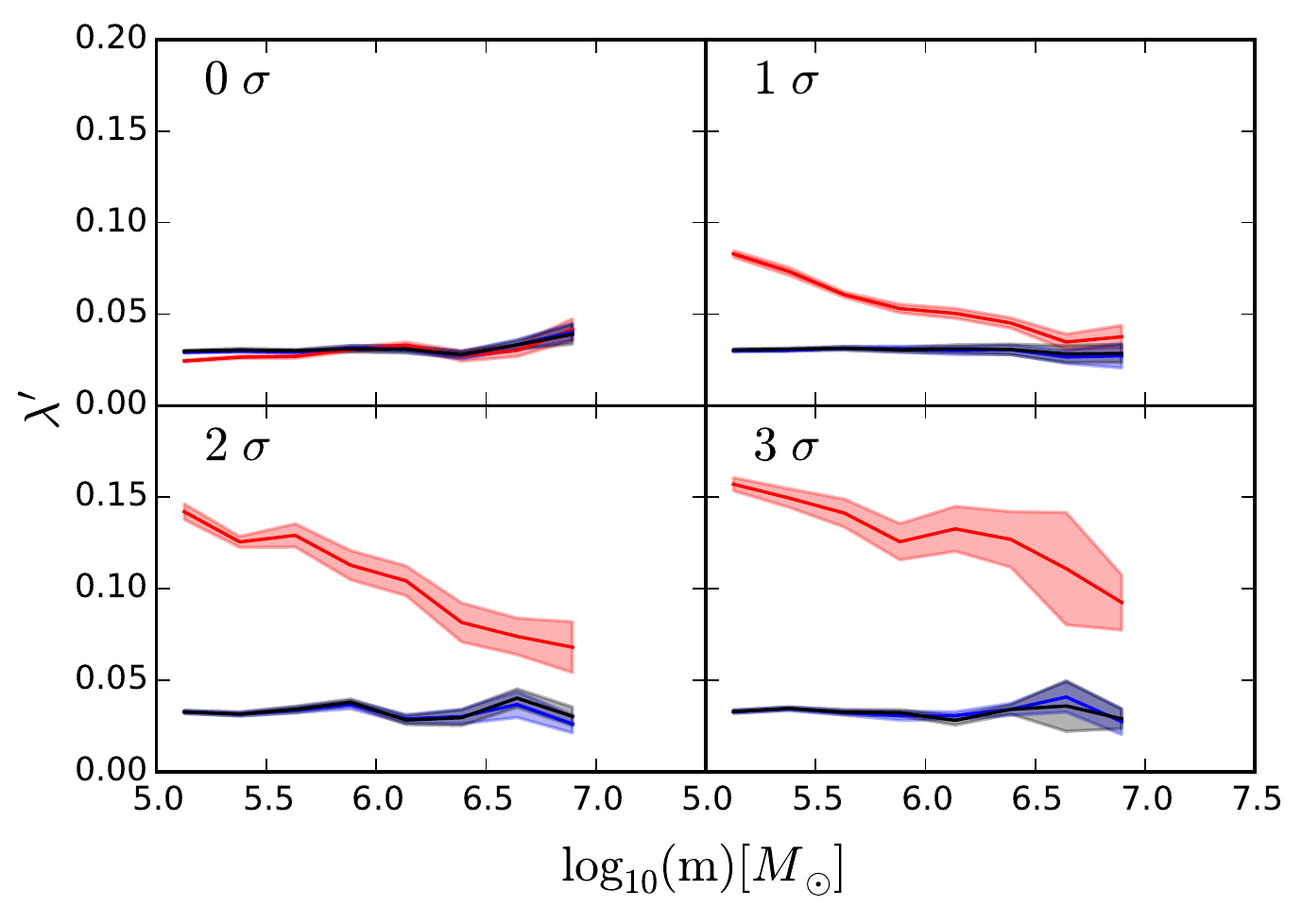}
\caption{The peak value of the spin parameter distribution, plotted as a function of halo mass, for the  gas (red), dark matter (black) and total matter (blue). The four panels show results for our four different simulations, as indicated in the top left corner of each panel.} 
\label{fig:Mass_vs_Lambda_AlleInEinemPlotMedian_Balken}
\end{figure}

On the other hand, when the streaming velocity is non-zero, we 
find a clear difference in behaviour. While for total matter the peak value of the spin parameter remains independent of halo mass for all streaming velocities, this is no longer true for the gas component. Instead, we see an increase in the peak value with decreasing halo mass. This can be seen more clearly in Figure \ref{fig:Mass_vs_Lambda_AlleInEinemPlotMedian_Balken}, where we plot only the peak value of the spin parameter distribution for each mass bin (the red and blue lines from Figure~\ref{fig:Mass_vs_Lambda_0Sig_013}), plus the same quantity for the dark matter (black lines). 

In the 1$\sigma$ streaming run, we see that $\lambda^{\prime}_{\rm gas} \simeq \lambda^{\prime}_{\rm tot}$ in the highest mass bin, but that it systematically increases above this value for decreasing halo mass, so that in the lowest mass bin it is almost three times as large. In the runs with even stronger streaming, we not only see a similar dependence on halo mass but also a clear offset between $\lambda^{\prime}_{\rm gas}$ and $\lambda^{\prime}_{\rm tot}$ even in the highest mass bin.

\begin{figure}
\includegraphics[width=0.99\columnwidth]{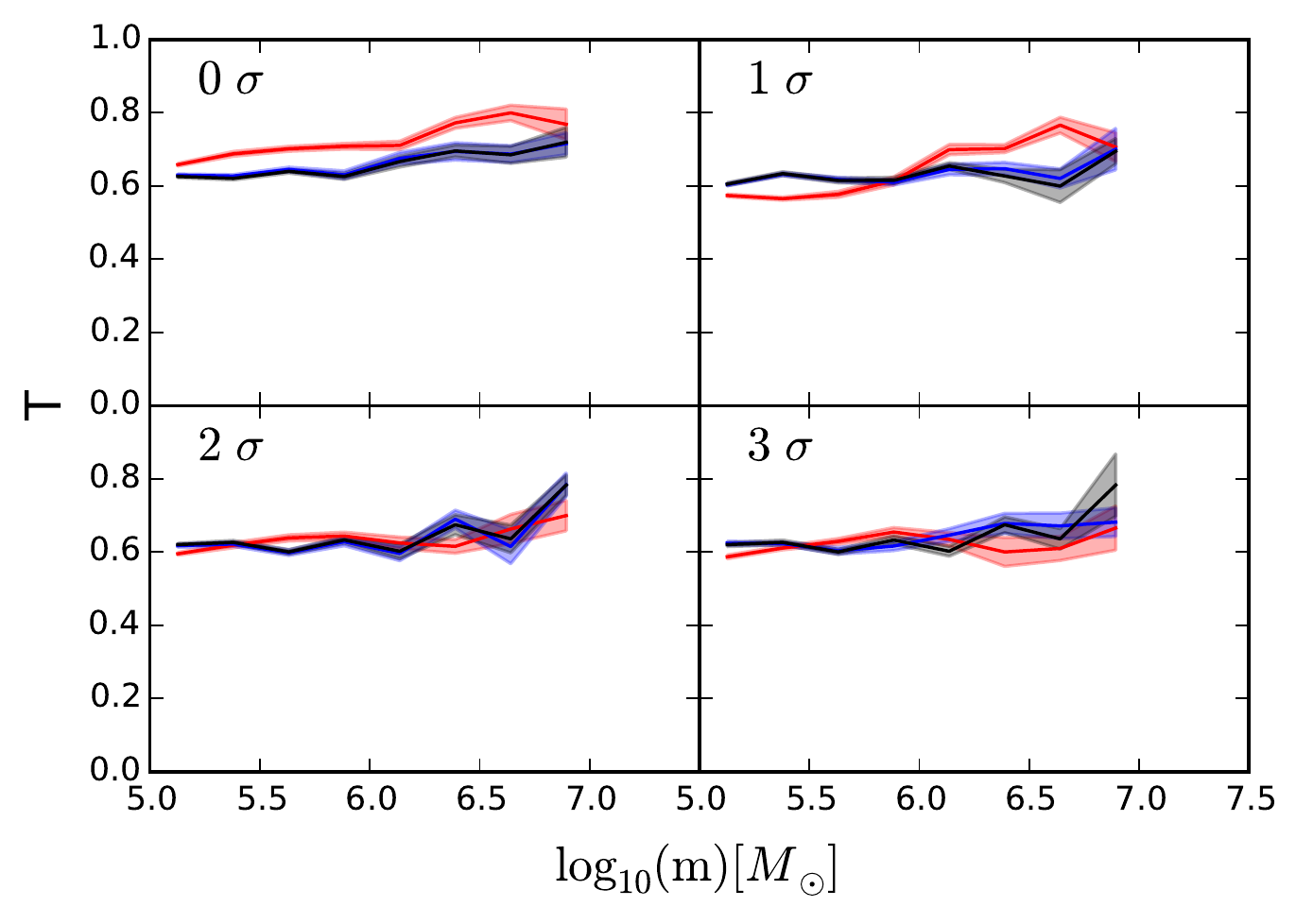}
\caption{Same as Figure \ref{fig:Mass_vs_Lambda_AlleInEinemPlotMedian_Balken}, but for the triaxiality of all haloes. The triaxiality is mostly independent of the halo mass, with gas values ranging from 0.6 to 0.8. There only is a very small dependence that higher mass halos are slightly more prolate than lower mass halos in the runs with small streaming velocity.} 
\label{fig:Mass_vs_Triax_AlleInEinemPlotMedian_Balken}
\end{figure}

\begin{figure}
\includegraphics[width=0.99\columnwidth]{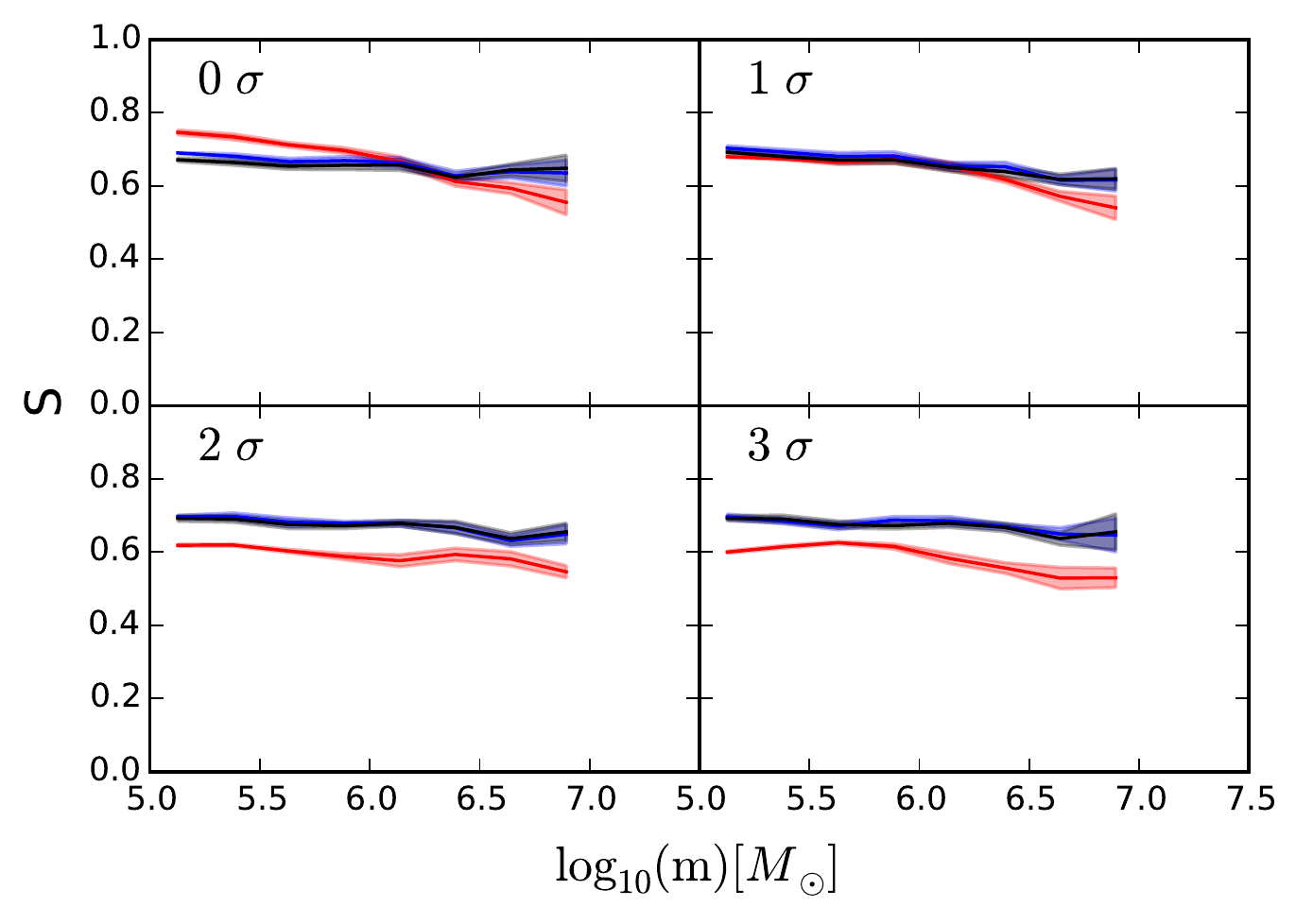}
\caption{Same as Figure \ref{fig:Mass_vs_Lambda_AlleInEinemPlotMedian_Balken}, but for the sphericity of all haloes. The values of the gas sphericity range from 0.5 to 0.8, with smaller halos being more spherical. This trend is stronger for the low streaming velocity runs. } 
\label{fig:Mass_vs_Sphaere_AlleInEinemPlotMedian_Balken}
\end{figure}

We have also investigated the mass dependence of the shape parameters (triaxiality $T{}$ and sphericity $S{}$), as shown in Figures \ref{fig:Mass_vs_Triax_AlleInEinemPlotMedian_Balken} and \ref{fig:Mass_vs_Sphaere_AlleInEinemPlotMedian_Balken}. In this case, the beta distribution is not always a good description of the distribution of $T$ and $S$ in each bin (especially in the low mass bins) and so the representative value we plot for each bin is the median value.

We see from the Figures that the triaxiality is largely independent of the halo mass, regardless of the streaming velocity, although there is a hint that in the low streaming runs, higher mass haloes are slightly more prolate than lower mass haloes. 
 
On the other hand, the sphericity does show a more pronounced mass dependence, with lower mass haloes being more spherical than higher mass haloes. When the streaming velocity is low, there is little difference between the sphericity of the gas distribution and that of the total mass, but when the streaming velocity is high there is a clear offset between the two, with the gas having a systematically less spherical distribution than the dark matter or the total mass.

\subsubsection{Radial dependence}
To further explore the impact of the streaming velocity on the spin parameter of the gas, we have also examined how this varies as a function of radius in the different simulations.
\begin{figure} 
\includegraphics[width=0.99\columnwidth]{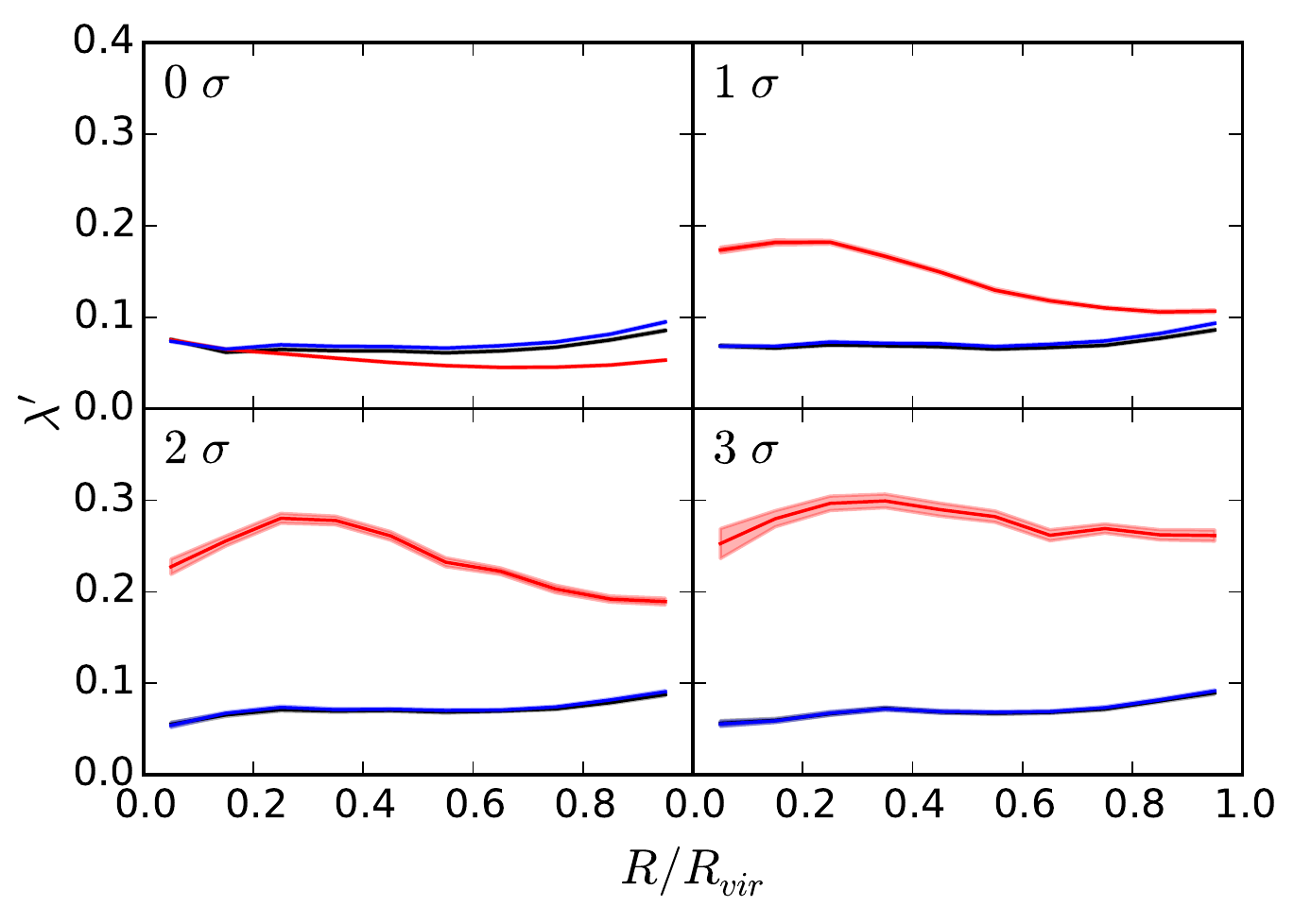}
\caption{Peak value of the spin parameter distribution as a function of radius (for shells with a width of $d=0.1 R_{\rm vir}$). We use the same color scheme as in Figure \ref{fig:Mass_vs_Lambda_AlleInEinemPlotMedian_Balken}, i.e.\ gas (red), dark matter (black) and total matter (blue).}
\label{fig:LambdaSlicesAllSigma}
\end{figure}
To do this, we split up each halo into ten separate shells of thickness 0.1 $R_\mathrm{Vir}$ and calculated the spin and shape parameters for each shell individually.\footnote{We chose this particular shell width to ensure that each shell would contain enough dark matter particles and gas cells to enable an accurate calculation of the spin and shape parameters.}

Comparison of the spins of shells with the same $R/R_{\rm Vir}$ in different haloes shows that the distribution can once again be represented as a log-normal. In Figure~\ref{fig:LambdaSlicesAllSigma} we show the peak value of this log-normal distribution as a function of $R/R_{\rm Vir}$ for our four different runs. The uncertainty in this peak value was calculated using the bootstrap method and is indicated by the shaded region. 
As before, the calculation was carried out with the various components and plotted in the usual colors gas (red), dark matter (black) and total matter (blue).

We can immediately see from Figure \ref{fig:LambdaSlicesAllSigma} that the spin parameter of the gas component changes significantly with the streaming velocity, while the distributions for dark matter and for the entire halo only change slightly. The distribution of $\lambda^{\prime}$ with radius for the dark matter remains fairly flat in all four simulations, with a slight rise to  $\lambda^{\prime}_\mathrm{dm} \sim 0.1$ in the outer third of the halo, and the distribution of $\lambda^{\prime}$ for the total mass behaves similarly. For the gas, we also find a flat distribution in the no streaming run. However, in the runs with streaming, we find qualitatively different behaviour. In this case, $\lambda^{\prime}$ is higher close to the centre of the halo than at the virial radius, and hence largely decreases with $R / R_{\rm Vir}$. However, the largest value of $\lambda^{\prime}$ is not found at the halo centre, but instead is offset to $R/R_{\rm Vir} \sim 0.2$--0.3. It is also clear that the degree to which $\lambda^{\prime}$ drops between this peak and the virial radius also depends on the streaming velocity: with a low streaming velocity, $\lambda^{\prime}$ decreases by almost a factor of two between $R = 0.3 R_{\rm Vir}$ and $R = R_{\rm Vir}$, while in the 3$\sigma$ streaming run, the drop is much smaller.
\begin{figure} 
\includegraphics[width=0.99\columnwidth]{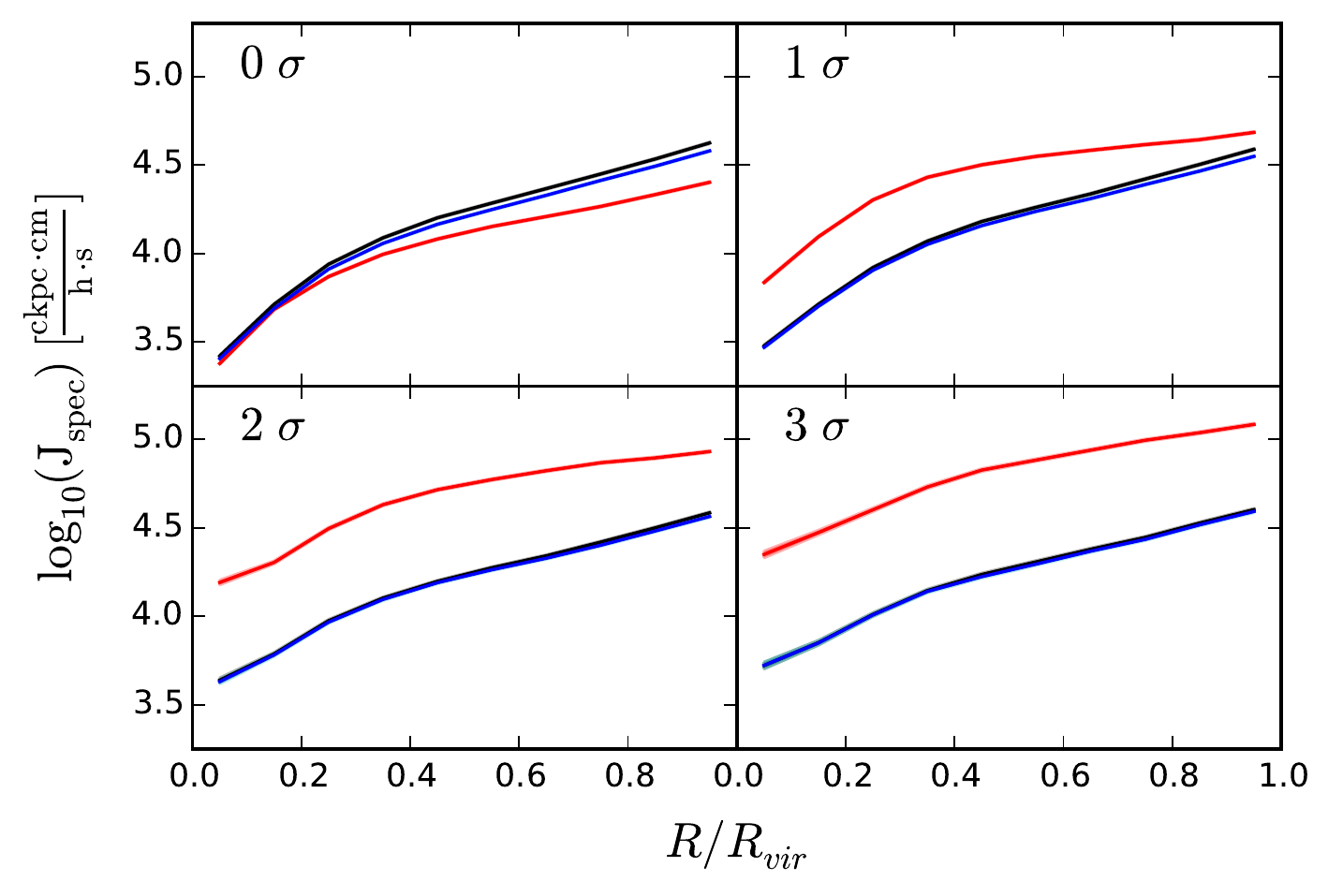}
\caption{Same as Figure \ref{fig:LambdaSlicesAllSigma}, but for the median value of the specific angular momentum of all haloes.}
\label{fig:JSlicesAllSigmaMedianBootstrap}
\end{figure}

To further explore this behaviour, we have also calculated the specific angular momentum $J_\mathrm{spec} = J / M{}$ as a function of radius for the different streaming velocities and for gas (red), dark matter (black) and total matter (blue). As the peak value, we choose the median of the specific angular momentum in each shell. 
The results are shown in Figure \ref{fig:JSlicesAllSigmaMedianBootstrap}. In every case, the specific angular momentum increases with increasing radius. As before, there is little change of the dark matter component and hence the combined halo when including streaming velocities. The gas component shows a higher specific angular momentum for the entire radius range for higher streaming velocities, consistent with Figure \ref{fig:LambdaSlicesAllSigma}. This is also in agreement with Figure~\ref{fig:PlotAlleSigmaNummer13}, where we see clumps of high density gas offset from the gravitational centre of the minihalo \citep[compare also][]{Chiou,Chiou19}. For minihaloes in the 1\,$\sigma{}$ streaming velocity simulation, the increase is stronger at smaller radii than at larger radii, which we interpret as a smaller offset of the denser gas from the centre than in the higher streaming velocity simulations. Further investigations have shown that in the run without streaming, any offset between the highest density gas and the halo centre is small (typically less than $0.1 R_{\rm vir}$; see Section~\ref{sect:toy} below), but that it increases significantly as we increase the streaming velocity. We therefore conclude that the 
higher spin parameter of the gas found in the higher streaming velocity simulations is not because the gas forms a larger or more rapidly rotating disk, but is instead due to the presence of dense clumps of gas at large distances from the halo centre with significant tangential velocities.

\subsection{Toy model} \label{sect:toy}
We construct a toy model to explain the increased spin value for minihaloes in high streaming velocity regions of the Universe. We have seen that the densest gas cell shifts out of the centre of the halo in regions of high streaming velocity. A gas clump rotating around the halo center can increase the angular momentum and hence the spin parameter of the gas component of this halo. For this model, we therefore split the $\lambda$-parameter into the internal component, which we assume to be equal to the spin parameter for 
no streaming velocity, $\lambda_\mathrm{int} = 0.0256$, and a clump component, $\lambda_\mathrm{clump}$. 
We rewrite the equation (3) for the gas component of the spin parameter:
\begin{eqnarray}
\lambda_\mathrm{gas} &=& \frac{|J_\mathrm{gas}|}{\sqrt{2} R_\mathrm{vir} M_\mathrm{gas} V_\mathrm{circ}} \\
&\approx& \lambda_\mathrm{int} + \frac{\int_\mathrm{r\,in\,clump}|\rho r v \mathrm{d}r^3|}{\sqrt{2} R_\mathrm{vir} M_\mathrm{gas} V_\mathrm{circ}}\\
&=& \lambda_\mathrm{int} + \frac{1}{\sqrt{2}}\frac{M_\mathrm{clump}}{M_\mathrm{gas}} \frac{R_\mathrm{clump}}{R_\mathrm{vir}} \frac{V_\mathrm{clump}}{V_\mathrm{circ}} .
\end{eqnarray}
We assume that the whole clump is moving with the same velocity and is very small, so that it can be approximated as a point mass with the distance $R_\mathrm{clump}{}$ to the halo centre, which we take as the separation between the densest gas cell and the centre of the halo. We further test a second approximation, where we assume that the clump has a tangential velocity equal to the Kepler velocity at the clump radius: 
\begin{equation}
v_\mathrm{clump} \approx v_\mathrm{Kepler}(R_\mathrm{clump}) = \sqrt{\frac{\mathrm{G} M_\mathrm{vir}}{R_\mathrm{clump}}}. 
\end{equation}
With this approximation, the ratio between the clump velocity and the circular velocity of the halo reduces to $\frac{V_\mathrm{clump}}{V_\mathrm{circ}} = \sqrt{R_\mathrm{vir}/{R_\mathrm{clump}}}$. This leads to the second approximation for $\lambda_\mathrm{clump}$: 
\begin{eqnarray}\label{equ:clump2}
\lambda_\mathrm{gas} &\approx& \lambda_\mathrm{int} + \frac{1}{\sqrt{2}}\frac{M_\mathrm{clump}}{M_\mathrm{gas}} \left(\frac{R_\mathrm{clump}}{R_\mathrm{vir}}\right)^{1/2} 
\end{eqnarray}

\begin{figure} 
\includegraphics[width=0.99\columnwidth]{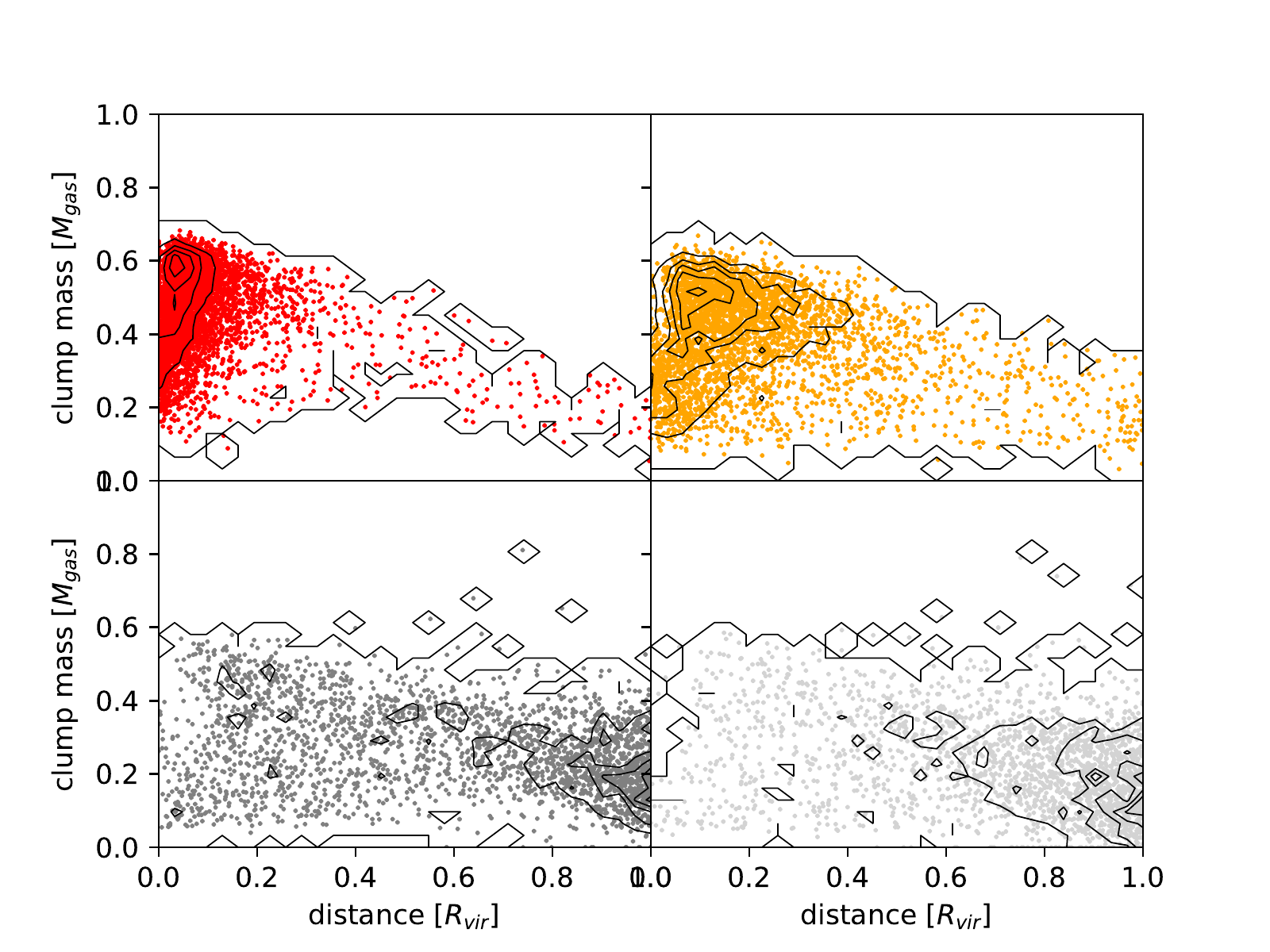}
\caption{Scatter plot of the gas clumps in the simulation, showing their masses (in units of the gas mass of the host halo) and distances from the halo centre (in units of the virial radius). Note that with the definition of clump we adopt here, there can be no more than one clump per halo, so each point in the plot corresponds to a different minihalo. From top left to bottom right the panels correspond to simulations with $0\sigma{}$ (red),  $1\sigma{}$ (orange),  $2\sigma{}$ (grey),  $3\sigma{}$ (light grey) streaming velocities.}
\label{fig:clump_abst}
\end{figure}
In a first step, we investigate whether haloes in streaming velocity regions systematically have dense clumps located outside of the halo center. For the purposes of this investigation, we define a clump to be a spherical region around the densest gas cell that contains all gas cells that have a higher density than the mean gas density of the respective halo and that are located no further than 10\% of the virial radius from the densest gas cell. Note that with this definition, we will identify a central density spike in a minihalo as a clump if that is where the densest gas cell is located. Note also that by definition we will identify only one clump per halo, since there is only one densest cell.

\begin{figure} 
\includegraphics[width=0.99\columnwidth]{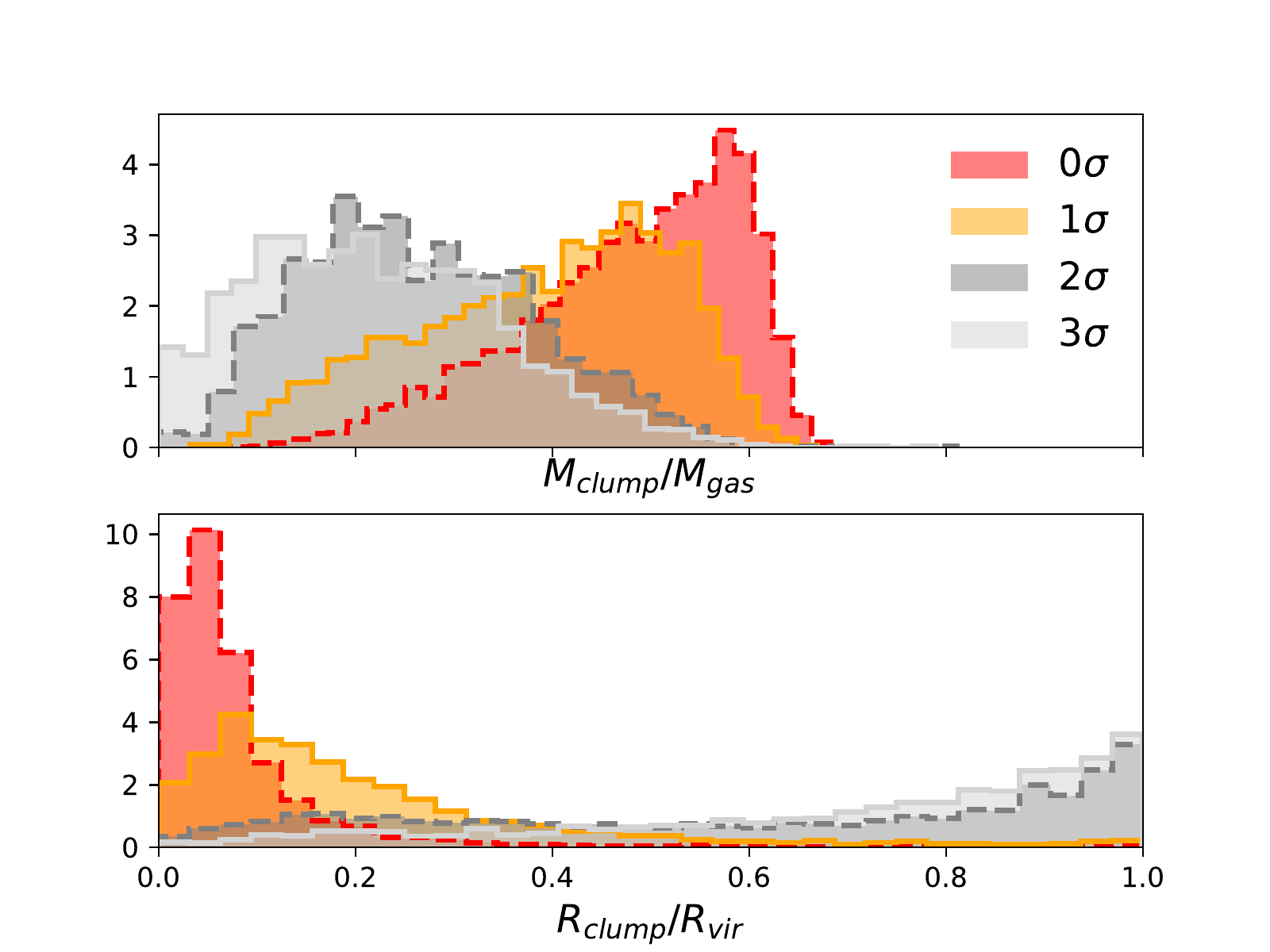}
\caption{Histograms of the gas clump properties. In the top panel, we show the distribution of clump masses in units of the total gas mass of the halo hosting the clump. In the bottom panel, we show the distribution of the distance of the gas clump to the halo center, in units of the virial radius. The streaming velocity simulations are colour coded: $0\sigma{}$ (red),  $1\sigma{}$ (orange),  $2\sigma{}$ (grey),  $3\sigma{}$ (light grey) streaming velocity. }
\label{fig:clump_abst2}
\end{figure}

In Figure \ref{fig:clump_abst}, we show the distribution of the clump masses and radii in the different simulations. To aid in the comparison of minihalos that cover a wide range of masses, we show clump masses in units of the total gas mass in the minihalo, $M_{\rm gas}$, and clump locations in units of the virial radius. For the $0\sigma{}$ (red) streaming velocity simulation, we find that most haloes have their 
dense gas at the halo centre, and the gas clumps are the most massive ones, containing 
50--60\% of the gas mass. In the $1\sigma{}$ (orange) streaming velocity simulation, the 
gas clumps are close to the halo centre, peaking at a distance of less than 10\% of the 
virial radius, with a tail extending to larger distances. These clumps are massive, too, 
with a distribution peaking at 50\% of the gas mass in the halo. 
This picture changes for higher streaming velocity regions (grey and light grey). 
The gas clumps generally contain less than 50\% of the gas in the halo. The distance between the dense gas clump and the halo centre is larger, with separations peaking at the maximum distance of the virial radius. This can be also seen in Figure \ref{fig:clump_abst2}, in which we show the one dimensional histograms 
of these clump properties.  

\begin{figure} 
\includegraphics[width=0.99\columnwidth]{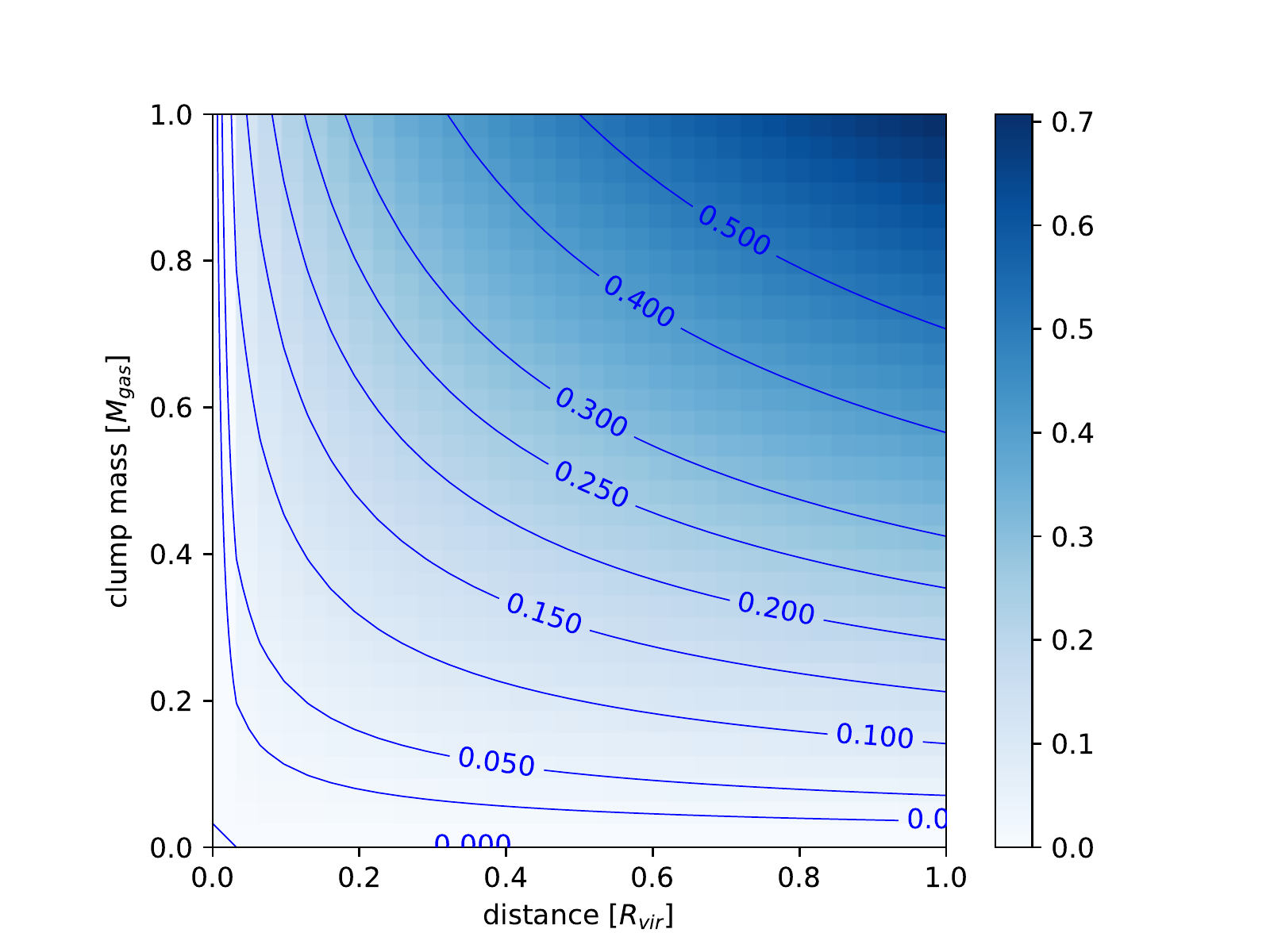}
\caption{Theoretical predictions of the additional spin parameter, $\lambda_\mathrm{clump}$, based on our simple model in Equation (\ref{equ:clump2}). }
\label{fig:clump_theory}
\end{figure}
\begin{figure} 
\includegraphics[width=0.99\columnwidth]{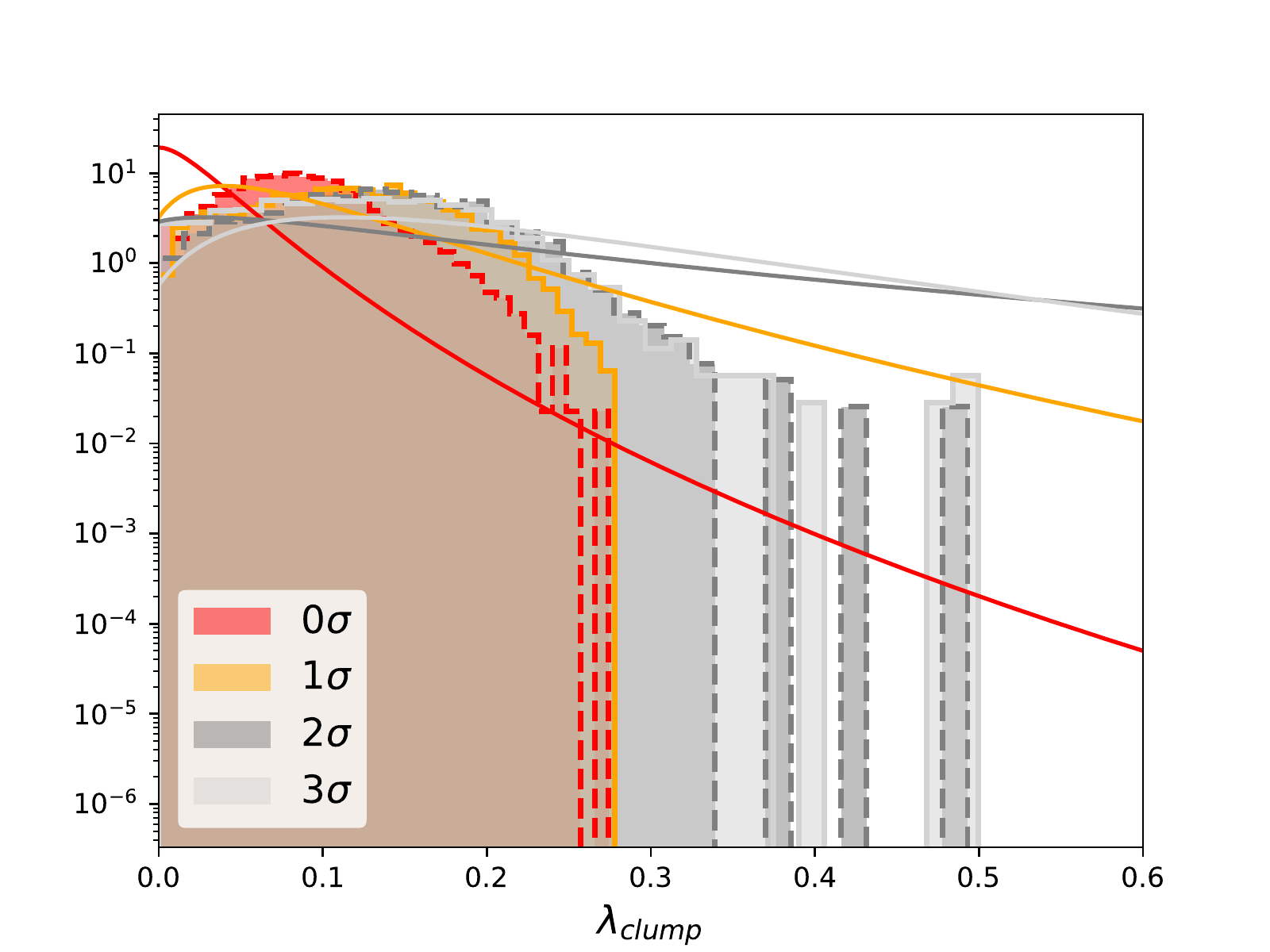}
\caption{Histogram of the additional spin parameter component, $\lambda_\mathrm{clump}$, using the Kepler velocity for the clump velocity with the approximation in Equation \ref{equ:clump2}. We use the same colour scheme as in Figure \ref{fig:clump_abst}, with $0\sigma{}$ (red),  $1\sigma{}$ (orange),  $2\sigma{}$ (grey), and  $3\sigma{}$ (light grey) streaming velocities. For comparison, we include the lognormal distributions that have been fitted to the spin parameter in Section \ref{Spin and shape properties} with solid lines. }
\label{fig:clump_simple}
\end{figure}
We can now estimate the contribution of the gas clump to to the total spin parameter. Following the approximation in Equation (\ref{equ:clump2}), we show the theoretical values of 
$\lambda_\mathrm{clump}$ in the 
$M_\mathrm{clump}/M_\mathrm{gas}$--$R_\mathrm{clump}/R_\mathrm{vir}{}$ plane in Figure \ref{fig:clump_theory}. One can see that in the bottom right of the figure (corresponding to a large distance and a 20-40\% clump mass), an additional spin parameter value of up to $\lambda_\mathrm{clump} = 0.1$--0.2 can be reached. This is the region that is most populated by 
haloes in the $3\sigma{}$ streaming velocity region.

The combination of this value with the internal spin value of $\lambda_\mathrm{int} = 0.0256$ yields total values similar to the peak values reached in the simulations with $2\sigma{}$ and $3\sigma{}$ streaming velocities, which are $\lambda_{2\sigma} = 0.1193{}$ and $\lambda_{3\sigma} = 0.1381{}$, respectively. The position and mass of the gas clump outside the halo center can therefore explain the increased spin parameter. This can be seen in Figure \ref{fig:clump_simple}, which shows a histogram of $\lambda_\mathrm{clump}$ values for each of the different streaming velocity simulations. However, the peak values of these distributions  overproduce the spin value of the $0\sigma{}$ and $1\sigma{}$ streaming velocity simulations. 

\begin{figure} 
\includegraphics[width=0.99\columnwidth]{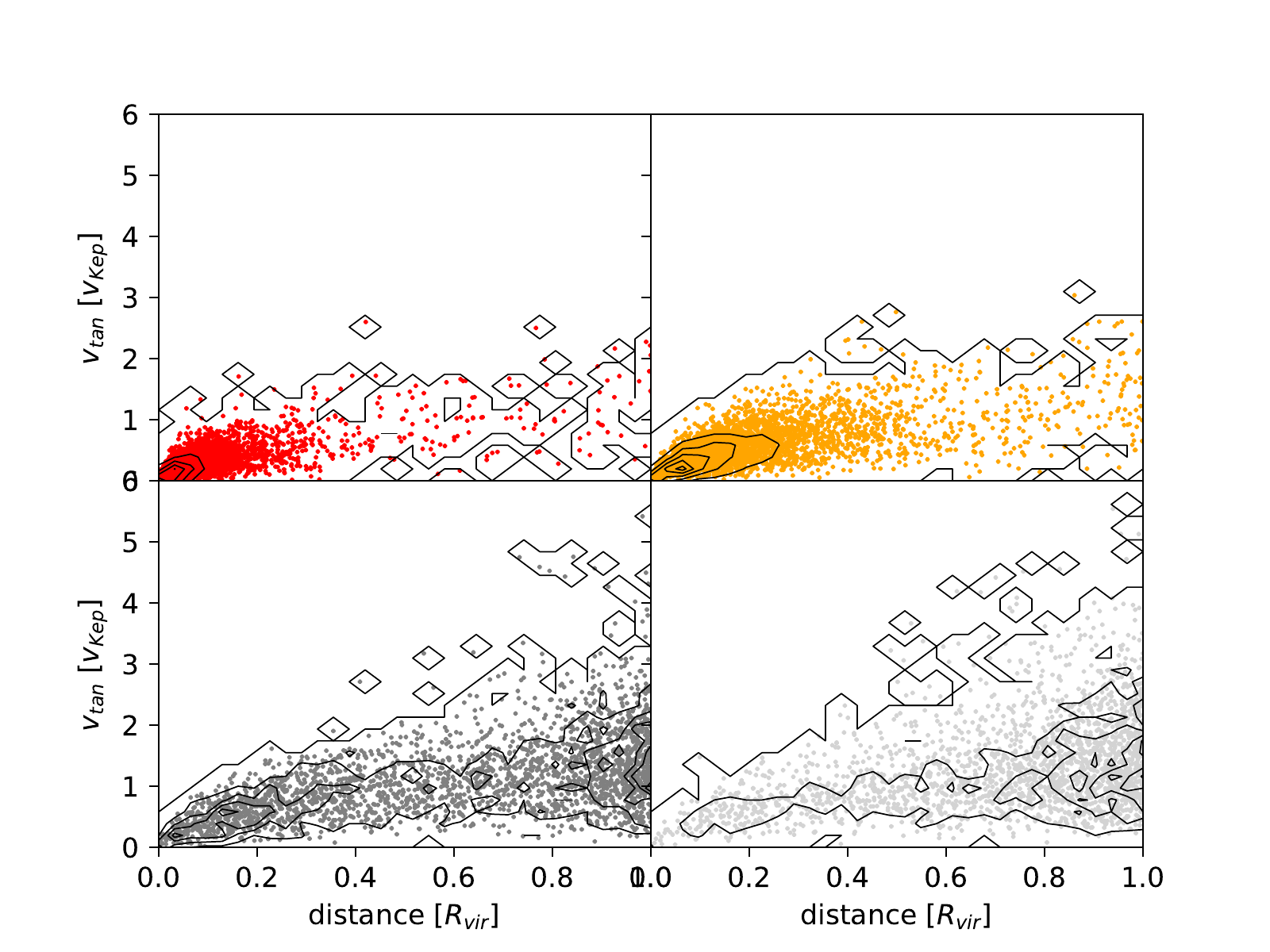}
\caption{Scatter plot of the gas clumps in the simulation. We show the tangential velocity (in units of the circular velocity of the halo) as a function of the clump distance (in units of the virial radius). We use the same colour scheme as in Figure \ref{fig:clump_abst}, with $0\sigma{}$ (red),  $1\sigma{}$ (orange),  $2\sigma{}$ (grey), and  $3\sigma{}$ (light grey) streaming velocities.}
\label{fig:clump_histvel}
\end{figure}

In a next step, we test our assumption that the clump roughly moves with the Kepler velocity at the clump radius. We compare the actual tangential velocity to the Kepler velocity at that radius and find that this assumption is approximately valid for gas clumps outside the halo centre. Figure \ref{fig:clump_histvel} shows the velocity ratios with respect to the clump distances. For higher streaming velocities, the  distribution gets broader, however, with more clumps moving faster than the Kepler speed. For the $0\sigma{}$ and $1\sigma{}$ streaming velocity simulations, the velocity is smaller than  Kepler, as most haloes have the dense gas in the halo centre. This explains why in these streaming velocity regions, the spin parameter peaks at smaller values than predicted by the simple toy model, as shown in Table \ref{Table:ValuesTabelle} and Figure \ref{fig:LambdaGASdmTOT0123Subplot}, respectively.

\begin{figure} 
\includegraphics[width=0.99\columnwidth]{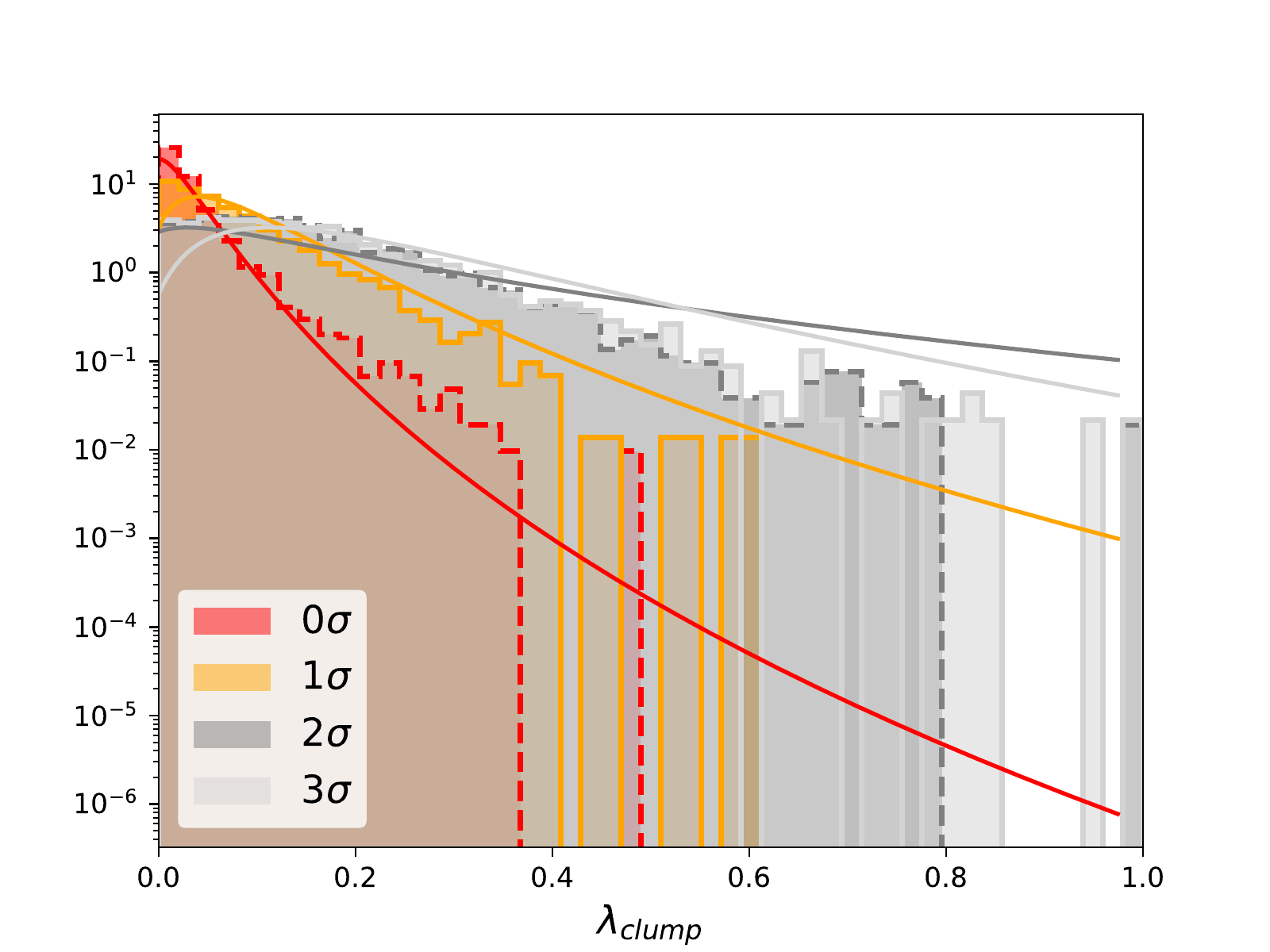}
\caption{Histogram of the additional spin parameter component, $\lambda_\mathrm{clump}$, using the measured tangential velocity of the clump. We use the same colour scheme as in Figure \ref{fig:clump_abst}.}
\label{fig:clump}
\end{figure}

In Figure \ref{fig:clump}, one can see that the additional clump parameter indeed grows and shifts to large value for high streaming velocity regions, while it clearly peaks at small values for the no-streaming velocity simulation. It matches the spin parameter distribution fitted in Section \ref{Spin and shape properties} and we can conclude that our toy model provides a good description of the increased spin parameter in regions of the Universe with increased streaming velocities.
\subsection{Angle Correlations} \label{Angle Correlations}

\begin{figure} 
\includegraphics[width=0.99\columnwidth]{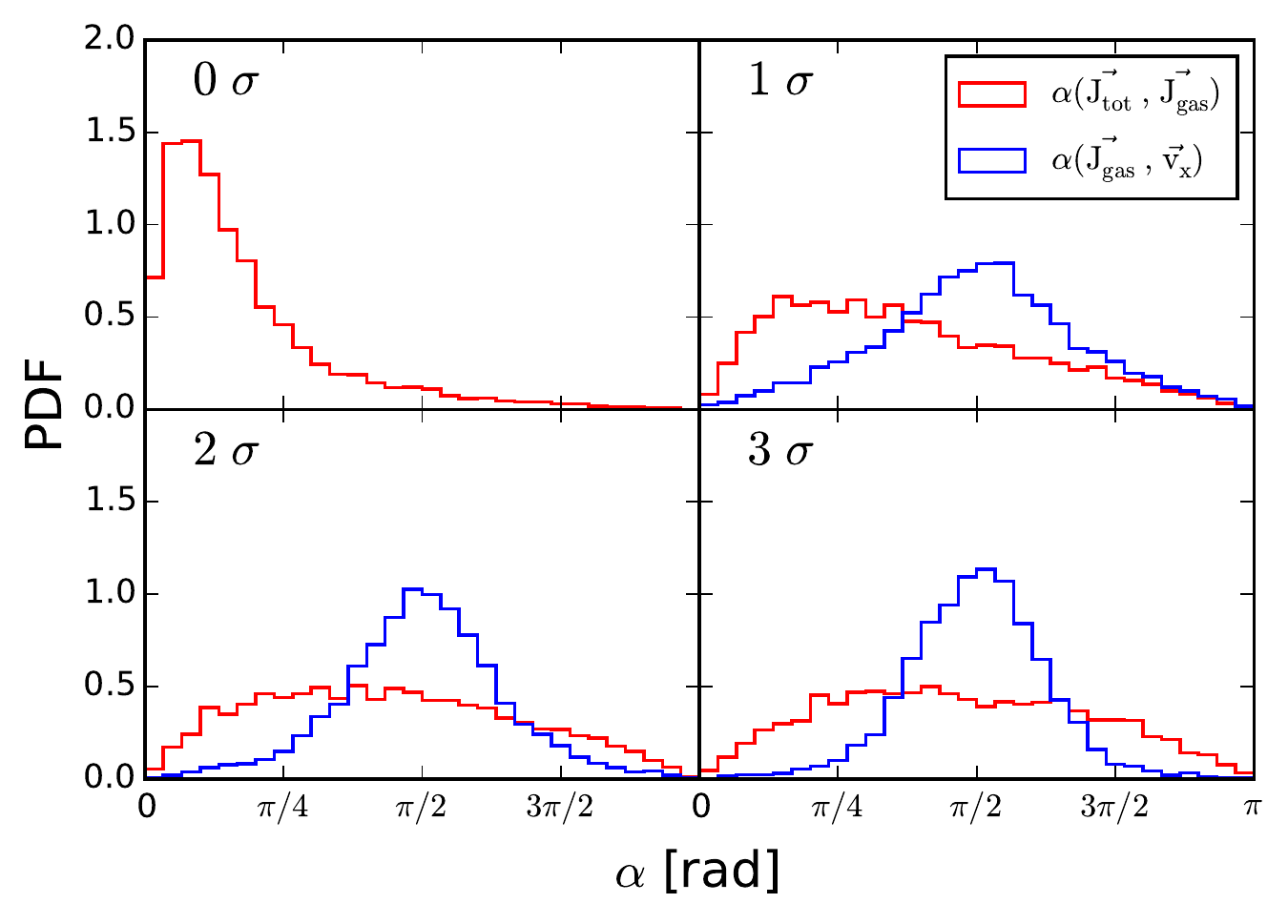}
\caption{Distribution of the angles between the gas and total halo components of the specific angular momentum (red) and between the specific gas angular momentum and the vector of the streaming velocity flow direction (blue). While the gas and total halo components are mostly correlated in the case of no streaming velocity, this correlation fades for the higher streaming velocity runs, being almost random for a streaming velocity of $3\sigma_\mathrm{rms}$. The gas angular momentum is mostly perpendicular to the streaming velocity direction. }
\label{fig:AnglecorrelationAlleInEinem2}
\end{figure}
From the previous section, it is clear that minihaloes formed in runs with higher streaming velocities tend to have gas components with higher spin parameters. We can gain some insight into why this is so if we examine how the angular momentum of the gas is oriented with respect to the angular momentum of the dark matter. In Figure~\ref{fig:AnglecorrelationAlleInEinem2}, we show the distribution of the angle $\alpha$ between the angular momentum vector for the total halo mass distribution and that for the gas.

In the run without streaming, we see that in most haloes the direction of the two vectors is highly correlated, with $\alpha$ peaking close to zero. The median angle is $\sim 23$ degrees, in good agreement with the value of around 30 degrees found in previous work \citep{Bosch,Liao}. However, as the streaming velocity increases, this correlation disappears: the distribution of $\alpha$ flattens and becomes consistent with a purely isotropic distribution \citep{Chiou}. We can see why this happens if we examine the alignment between the direction of the gas angular momentum vector and the direction of the streaming velocity in these runs (blue curves in Figure~\ref{fig:AnglecorrelationAlleInEinem2}). This peaks around an angle of 90 degrees, with this peak becoming increasingly pronounced as the streaming velocity increases. 

If the motion of the gas were purely due to the streaming, we would expect to recover a perfect 90 degree alignment between the streaming direction and the angular momentum (since $\vec{J} = \vec{r} \times \vec{v}$). In reality, the gas also has motions due to its infall into the dark matter potential well and due to the tidal torque acting on it from the surrounding distribution of matter. However, the fact that we nevertheless recover a clear peak in the alignment at 90 degrees in the runs with streaming demonstrates that in the majority of haloes it is the streaming that dominates the large-scale motion of the gas within the virial radius, particularly in the runs with high sigma streaming. Moreover, since the angular momentum of the dark matter is uncorrelated with the streaming, its alignment is random with respect to the streaming direction and hence also with respect to the gas in the high sigma streaming runs. On the other hand, in the run with no streaming, tidal torques dominate and so we recover a good correlation between the directions of the gas and dark matter angular momentum vectors since the same torques act on both components.

\subsection{Dense gas} \label{cold}
\begin{figure*} 
\includegraphics[width=1.99\columnwidth]{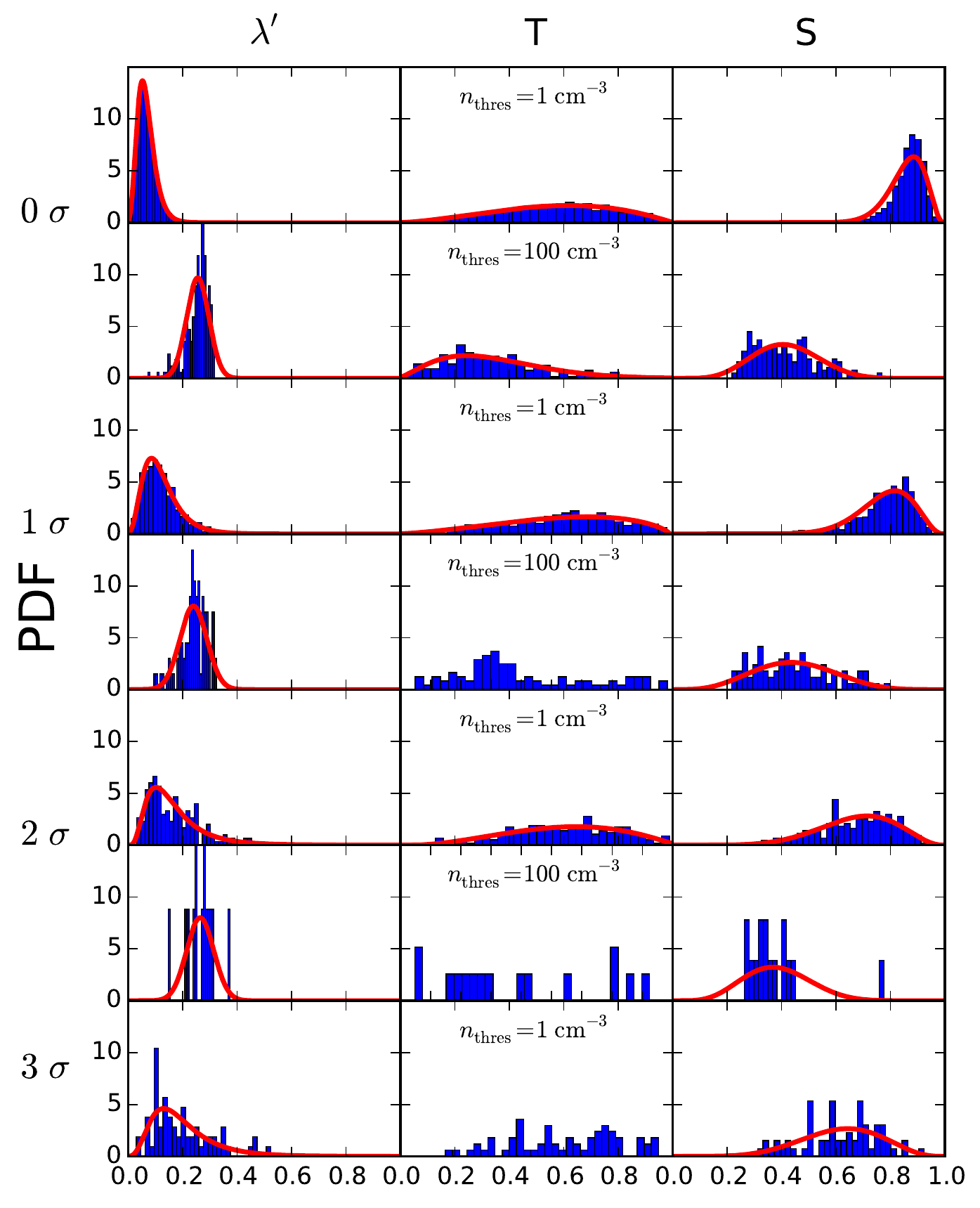}
\caption{Spin and shape distributions for cold dense gas haloes at 0, 1, 2 and 3$ \sigma$ streaming. Within the haloes only gas cells up to a certain radius are considered. This radius corresponds to the distance of the most distant gas cell with a density above the specified threshold density $n_{\rm thres}$. Where possible, we include the fitted distributions (log-norm and beta) as red lines in the figure. However, in some cases there is too little data to allow for a meaningful fit and in these cases no line is shown. No values are shown for the 3$\sigma$ run and $n_{\rm thres} = 100 \: {\rm cm^{-3}}$ as none of the haloes in this run have enough gas cells at densities above this threshold to allow meaningful values of $\lambda^{\prime}$, $T$ or $S$ to be computed.}
\label{fig:LaTrSpCgasAllpart}
\end{figure*}

It is also interesting to examine the impact of streaming on the spin of the cold dense gas found at the centre of the subset of minihaloes that are capable of forming stars. In Paper I, we showed that in the absence of streaming, there is no correlation between the spin of this gas and that of the halo as a whole, implying that the latter cannot be used to predict the former, contrary to previous conjectures in the literature \citep[e.g.][]{Souza}. Does this result still hold in runs that include the effects of streaming?

To investigate this, we have examined the behaviour of gas above two different density thresholds, $n_{\rm thres} = 1 \, {\rm cm^{-3}}$ and $n_{\rm thres} = 100 \, {\rm cm^{-3}}$. In each halo, we first calculate the distance from the centre 
of the halo 
(defined by the potential minimum of the halo)
to the farthest cell with a density above $n_{\rm thres}$. We then calculate the spin, sphericity and triaxiality of the gas contained within a sphere with a radius equal to this distance. 
In order to allow us to make a meaningful statement about the spin and the shape of the haloes, we require a minimum number of 25 cold dense cells per halo. Haloes that do not satisfy this requirement are not included in the analysis. Since streaming hampers gas cooling, particularly in low mass haloes \citep[see e.g.][]{schauer18}, the number of haloes considered here decreases for high streaming velocities. The number of haloes included in the analysis for each combination of simulation and $n_{\rm thres}$ is summarized in Table~\ref{Table:Anzahl}. Note that in the 3$\sigma$ streaming simulation, there are no haloes with 25 gas cells denser than $n_{\rm thres} = 100 \: {\rm cm^{-3}}$, preventing us from analyzing the properties of the dense gas in this case. 

In Figure \ref{fig:LaTrSpCgasAllpart}, the histograms (blue) and their distributions (red) are shown for both the spin and the shape of the dense gas in our four simulations. The different columns show $\lambda^{\prime}$, $T$ and $S$, respectively, for simulations with no streaming (first two rows), 1$\sigma$ streaming (third and fourth rows), 2$\sigma$ streaming (fifth and sixth rows) and 3$\sigma$ streaming (last row). The value of $n_{\rm thres}$ considered in each case is indicated in the plot. In the simulations with higher streaming velocities, the spin parameter, triaxiality and sphericity histograms are no longer well fit by log-normal or beta distributions, respectively, particularly when $n_{\rm thres} = 100 \, {\rm cm^{-3}}$. This is likely a consequence of the small number of haloes we are dealing with in these cases. As a result, we cannot easily calculate meaningful peak values for all of the histograms. Nevertheless, some basic trends are clear.

\begin{table}
\centering
\begin{tabular}{|c|cccc}
$n_\mathrm{thres}\ [\mathrm{cm}^{-3}]$       & 0$ \sigma$     & 1$ \sigma$   & 2$ \sigma$    & 3$ \sigma$  \\ \hline
1    &           2604 & 831          & 199           & 64 \\ 
100  &           206  & 87           & 15            & - 
\end{tabular}
\caption{Number of haloes in each simulation with at least 25 gas cells with densities greater than the specified threshold density.}
\label{Table:Anzahl}
\end{table}

Figure~\ref{fig:LaTrSpCgasAllpart} demonstrates that in gas denser than $n_{\rm thres} = 1 \: {\rm cm^{-3}}$ (roughly a factor of ten higher than the virial density at this redshift), we recover a very similar result to the one we found in Section~\ref{Spin and shape properties} for the total gas content, namely that as the streaming velocity increases, so does the spin parameter, while the sphericity decreases. However, if we turn our attention to gas denser than $n_{\rm thres} = 100 \: {\rm cm^{-3}}$, we see that in this case, neither the spin parameter nor the shape of gas distribution show any clear dependence on the streaming velocity. 
 
\begin{figure} 
\includegraphics[width=0.99\columnwidth]{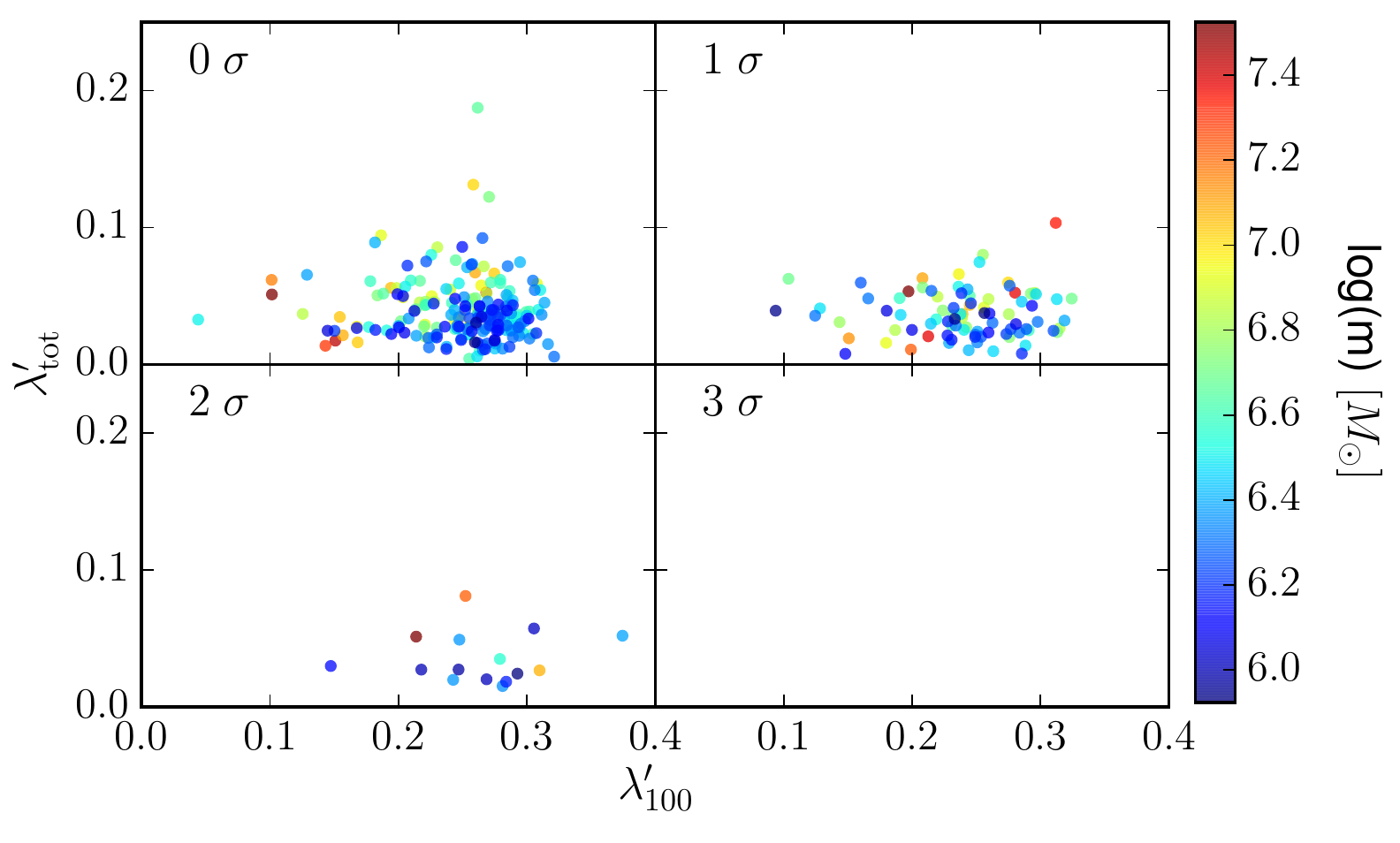}
\caption{Total spin parameter ($\lambda^{\prime}_{\rm tot}$) and spin parameter of gas denser than $n_{\rm thres} = 100 \: {\rm cm^{-3}}$ ($\lambda^{\prime}_{100}$), shown for each halo with at least 25 gas cells denser than $n_{\rm thres}$ and for the four different runs. The points are color-coded by the total mas of the halo. In the run with $3 \sigma$ streaming, there are no minihaloes with 25 cells above the density threshold.}
\label{fig:CrossLambda}
\end{figure}

We have also examined whether there is any correlation between the spin parameter of the dense gas and that for the halo as a whole (Figure~\ref{fig:CrossLambda}). In Paper I, we showed that in the absence of streaming, these two quantities are not correlated. Figure~\ref{fig:CrossLambda} demonstrates that this important result continues to hold in runs that include streaming.

Both of these results suggest that the spin and shape of the dense, gravitationally-collapsing gas are determined primarily by the details of the collapse itself, and preserve little or no memory of the state of the gas on large scales. This also implies that the properties of the stars that ultimately form from this gas are unlikely to be strongly affected by the value of the streaming velocity.

\section{Conclusion}\label{Conclusion}
In this paper, we have investigated the spin and shape distributions of a large sample of minihaloes formed in simulations with streaming velocities ranging from zero to $3\sigma$. We examine the state of the simulations at a redshift of $z = 14$ and only consider haloes with a minimum mass of at least $\mathrm{M_{min}} = 10^{5}\ \mathrm{M_{\odot}}$. This results in 7982 haloes for the simulation with no streaming, 6282 haloes for the case of 1$ \sigma$ streaming, 4798 haloes for 2$ \sigma$ streaming and 4263 haloes for 3$ \sigma$ streaming. As well as measuring the spin and shape distributions for the full sample of minihaloes, we have also explored how these properties vary as a function of halo mass. In a subset of the full minihalo sample, gas cools and undergoes runaway gravitational collapse. In these haloes, we have quantified the spin and shape distributions of the dense gas and have examined whether the spin and shape of the dense gas are correlated with the same properties measured on the scale of the halo as a whole. Below, we summarize our main results.

\begin{itemize}
\item Streaming velocities only affect the spin and shape distributions of the gas component in the minihaloes. Their effect on the dark matter component is negligibly small.

\item As the streaming velocity increases, the spin parameter of the gas component increases. The gas component of the halo is less spherical and less prolate for a non-zero streaming velocity than for the case of no streaming velocity \citep[compare][]{Druschke}.

\item The spin parameter of minihaloes in a region of the Universe with no streaming velocity is independent of mass. However, the minihalo shape has a slight dependence on mass: more massive minihaloes tend to be slightly more prolate and less spherical. 

\item In regions with streaming, the spin parameter of the gas in the minihaloes becomes mass dependent. Low-mass minihaloes develop higher spin parameters than higher-mass minihaloes. The shape
parameters, on the other hand, become completely independent of mass. The effect on dark matter is once again negligible.

\item The center of rotation of the halo and the position of the most bound particle can deviate significantly from each other under the influence of streaming velocities. This can lead to an increase of the spin parameter outside the centre of the minihalo. Figure \ref{fig:PlotAlleSigmaNummer13} shows an example of a rotational offset that is very strong for a streaming velocity of 3$ \sigma$. More quantitatively, Figure \ref{fig:JSlicesAllSigmaMedianBootstrap} shows that the specific angular momentum of the gas component increases with increasing streaming velocities at all radii. However, for a streaming velocity of 1\,$\sigma$, the change is more significant at the centre than for higher streaming velocity simulations, showing that the gas clumps causing the large spin move to larger radii.

\item We can explain the increased gas spin parameter in regions of non-zero streaming velocity with a simple toy model. For that, we split the spin parameter into an intrinsic (corresponding to no streaming velocity) value and an additional clump component. If we assume that the additional angular momentum is caused by a single gas clump that orbits the halo centre, we can approximately reproduce the shape of the spin parameter distribution (compare Figures \ref{fig:clump_theory} and  \ref{fig:clump_histvel}). The velocity of that gas clump is similar to the Kepler velocity expected at that radius in the case of a streaming velocity of 2--3$\sigma$, but on average lower for no streaming velocity or a moderate streaming velocity of 1$\sigma$. 

\item The streaming velocity also affects the orientation the angular momentum of the gas component. As the streaming velocity increases, it becomes aligned increasingly strongly in a direction perpendicular to the streaming motion. Since the dark matter haloes themselves have angular momenta that are aligned randomly with respect to the streaming motion, the result is that the alignment between the gas and dark matter angular momentum in any given minihalo becomes increasingly random.

\item The spin and shape distributions of dense, gravitationally collapsing gas within the minihaloes are uncorrelated with the values on the scale of the virial radius and unaffected by the strength of the streaming velocity. The value of the streaming velocity is therefore unlikely to directly affect the properties of the stars that form from this collapsing gas. As the minimum halo mass for Pop~III star formation increases with streaming velocity, we expect a globally reduced and delayed star formation rate in regions of high streaming velocity \citep[compare e.g. with][]{schauer18}
\end{itemize}

\section*{Acknowledgments}
The authors would like to thank Volker Bromm, Mattis Magg, Smadar Naoz and Naoki Yoshida for fruitful discussions. They would also like to thank the anonymous referees for constructive reports that helped to improve the paper. 
Support for this work was provided by NASA through the NASA Hubble Fellowship grant HST-HF2-51418.001-A awarded  by  the  Space  Telescope  Science  Institute,  which  is  operated  by  the Association  of  Universities for  Research  in  Astronomy,  Inc.,  for  NASA,  under  contract NAS5-26555.  
The authors acknowledge support from the European Research Council under the European Community's Seventh Framework Programme (FP7/2007 - 2013) via the ERC Advanced Grant ``STARLIGHT: Formation of the First Stars" (project number 339177). SCOG and RSK also appreciate support from the Deutsche Forschungsgemeinschaft (DFG, German Research Foundation) -- Project-ID 138713538 -- SFB 881 (``The Milky Way System'', subprojects B1, B2 and B8) and SPP 1573 , ``Physics of the Interstellar Medium'' (grant number GL 668/2-1). 
SCOG and RSK further acknowledge support from the DFG through Germany's Excellence Strategy EXC-2181/1 - 390900948 (the Heidelberg STRUCTURES Excellence Cluster).
The authors gratefully acknowledge the Gauss Centre for Supercomputing e.V.\ (www.gauss-centre.eu) for providing computing time on the GCS Supercomputer SuperMUC at Leibniz Supercomputing Centre. 
The authors also acknowledge support by
the state of Baden-W\"urttemberg through bwHPC and the German
Research Foundation (DFG) through grant INST 35/1134-1 FUGG. 
The data underlying this article will be shared on reasonable request to the corresponding author. 
The authors would like to thank the Cox Fund for providing a travel grant which lead to the completion of this manuscript. 
\setlength{\bibhang}{2.0em}
\setlength\labelwidth{0.0em}
\bibliographystyle{mn2e}
\bibliography{sample}
\label{lastpage}

\end{document}